\documentclass[12pt,preprint]{aastex}
\usepackage{natbib}
\slugcomment{Submitted to Astrophysical Journal Supplement }

\shorttitle{Fermi LAT Bright Source List}
\shortauthors{Abdo et al.}

\begin{document}


\title{Fermi Large Area Telescope Bright Gamma-ray Source List}


\author{The {\it Fermi} LAT Collaboration\\ 
A.~A.~Abdo\altaffilmark{2,3}, 
M.~Ackermann\altaffilmark{4}, 
M.~Ajello\altaffilmark{4}, 
W.~B.~Atwood\altaffilmark{5}, 
M.~Axelsson\altaffilmark{6,7}, 
L.~Baldini\altaffilmark{8}, 
J.~Ballet\altaffilmark{1,9}, 
D.~L.~Band\altaffilmark{10,11,12}, 
G.~Barbiellini\altaffilmark{13,14}, 
D.~Bastieri\altaffilmark{15,16}
% ...
}
\altaffiltext{1}{Corresponding authors: J.~Ballet, jean.ballet@cea.fr; S.~W.~Digel, digel@stanford.edu; I.~A.~Grenier, isabelle.grenier@cea.fr; D.~J.~Thompson, David.J.Thompson@nasa.gov.}
\altaffiltext{2}{National Research Council Research Associate}
\altaffiltext{3}{Space Science Division, Naval Research Laboratory, Washington, DC 20375}
\altaffiltext{4}{W. W. Hansen Experimental Physics Laboratory, Kavli Institute for Particle Astrophysics and Cosmology, Department of Physics and SLAC National Accelerator Laboratory, Stanford University, Stanford, CA 94305}
\altaffiltext{5}{Santa Cruz Institute for Particle Physics, Department of Physics and Department of Astronomy and Astrophysics, University of California at Santa Cruz, Santa Cruz, CA 95064}
\altaffiltext{6}{The Oskar Klein Centre for Cosmo Particle Physics, AlbaNova, SE-106 91 Stockholm, Sweden}
\altaffiltext{7}{Department of Astronomy, Stockholm University, SE-106 91 Stockholm, Sweden}
\altaffiltext{8}{Istituto Nazionale di Fisica Nucleare, Sezione di Pisa, I-56127 Pisa, Italy}
\altaffiltext{9}{Laboratoire AIM, CEA-IRFU/CNRS/Universit\'e Paris Diderot, Service d'Astrophysique, CEA Saclay, 91191 Gif sur Yvette, France}
\altaffiltext{10}{Center for Research and Exploration in Space Science and Technology (CRESST), NASA Goddard Space Flight Center, Greenbelt, MD 20771}
\altaffiltext{11}{Deceased}
\altaffiltext{12}{NASA Goddard Space Flight Center, Greenbelt, MD 20771}
\altaffiltext{13}{Istituto Nazionale di Fisica Nucleare, Sezione di Trieste, I-34127 Trieste, Italy}
\altaffiltext{14}{Dipartimento di Fisica, Universit\`a di Trieste, I-34127 Trieste, Italy}
\altaffiltext{15}{Istituto Nazionale di Fisica Nucleare, Sezione di Padova, I-35131 Padova, Italy}
\altaffiltext{16}{Dipartimento di Fisica ``G. Galilei", Universit\`a di Padova, I-35131 Padova, Italy}
% ...


\begin{abstract}
Following its launch in June 2008, the {\it Fermi Gamma-ray Space Telescope (Fermi)} began  a sky survey in August.  The Large Area Telescope (LAT) on {\it Fermi} in 3 months produced a deeper and better-resolved map of the $\gamma$-ray sky than any previous space mission.  We present here initial results for energies above 100 MeV for the 205 most significant (statistical significance greater than $\sim$10-$\sigma$) $\gamma$-ray  sources in these data.  These are the best-characterized and best-localized {\bf point-like (i.e., spatially unresolved)} $\gamma$-ray sources in the early-mission data.


\end{abstract}


\keywords{ Gamma rays: observations --- surveys --- catalogs; Fermi Gamma-ray Space Telescope; PACS: 95.85.Pw, 98.70.Rz}

\section{Introduction}

Collections of information about what can be seen in the sky range from simple lists to complex catalogs.  For high-energy $\gamma$-rays (photon energies above 100~MeV), the first effort of this type was a COS-B source list \citep{Hermsen77}, followed by the second COS-B catalog \citep{COSB2}.  The Energetic Gamma Ray Experiment Telescope (EGRET) on the {\it Compton Gamma Ray Observatory} yielded several catalogs, culminating in the third EGRET Catalog \citep[3EG;][]{3rdCat} and an alternate catalog, EGR \citep{EGR}, but also including a catalog of just the sources seen above 1~GeV (Lamb and Macomb, 1997).  The AGILE telescope has recently released its first catalog  \citep{AGILE}\footnote{See http://www.asdc.asi.it/agilebrightcat/}.  The rapidly-changing field of TeV $\gamma$-ray astronomy has a number of on-line catalogs, e.g., TeVCat\footnote{http://tevcat.uchicago.edu/}, {\bf  a frequently-updated compilation of announced TeV sources from ground-based observatories.} 

The {\it Fermi Gamma-ray Space Telescope (Fermi)} Large Area Telescope (LAT) is a successor to EGRET, with greatly  improved sensitivity, {\bf angular } resolution, and energy range.  This paper presents a list of bright LAT sources that
have statistical significances of 10-$\sigma$ or higher, based on the
first three months of survey data. Although the first official LAT catalog is planned for release after the first year of operations, {\bf (after the LAT gamma-ray data themselves become publicly available)\footnote{See http://fermi.gsfc.nasa.gov/ssc/proposals/.}, this early list of bright sources was released to enable multiwavelength studies by the broader community and to support proposal preparation for Cycle 2 of the Fermi Guest investigator program. }

The reader is cautioned to avoid generalizing from this sample of sources.  Some particular features are:

\begin{itemize}
  \item The source list is not a complete summary of sources seen by the LAT.  Many additional sources are detected with lower confidence levels in the LAT data than are included here (Sect.~\ref{run:TS}); 
    
  \item The source list is not flux limited and hence not uniform.  Only sources above a 10-$\sigma$  statistical significance are included, as described below. {\bf Moreover,} owing to the strong energy dependence both of the angular resolution of the LAT and of the intensities of backgrounds, the limiting flux is dependent on spectral hardness. 
 Because $\gamma$-ray sources are seen against a background of diffuse gamma radiation, which is highly non-uniform across the sky, e.g., \citep{Hunter97,SMR}, the limiting flux for a given statistical significance {\bf and spectral shape} varies with position (Sect.~\ref{run:TS}). 
 
 \item The source list does not include detailed information about the energy spectra of individual sources. 
  
 \end{itemize}

Because this list is a step toward the first LAT catalog, we adopt the terminology for sources that will be used in that catalog, with a 0 prefix.  The source designation is \texttt{0FGL JHHMM.m+DDMM} where the \texttt{0} refers to the preliminary nature of this list and \texttt{FGL} represents {\it Fermi} Gamma-ray LAT  (Sect.~\ref{run:sourcelist}). 

\section{Gamma-ray Detection with the Large Area Telescope}
\label{run:LAT}

The LAT is a pair-production telescope \citep{LATpaper}. The tracking section has 36 layers of silicon microstrip detectors to record the tracks of charged particles, interleaved with 16 layers of tungsten foil (12 thin layers, 0.03 radiation length, at the top or front of the instrument, followed by 4 thick layers, 0.18 radiation length, in the back section) to promote $\gamma$-ray pair conversion.
 {\bf Below the tracker lies} an array of CsI crystals to determine the 
$\gamma$-ray energy. The tracker is surrounded by segmented charged-particle {\bf anticoincidence } detectors (plastic scintillators 
with photomultiplier tubes) to reject cosmic-ray backgrounds. The LAT's improved sensitivity 
compared to EGRET stems from a large peak effective area ($\sim$8000 cm$^2$, or $\sim$6 times greater than 
EGRET's), large field of view ($\sim$2.4 sr, or nearly 5 times greater than EGRET's), good background rejection, superior angular resolution (68\% containment angle $\sim0.6^\circ$ at 1 
GeV for the front section and about a factor of 2 larger for the back section,{\bf vs. $\sim$1.7$^\circ$ at 1 GeV for EGRET; \citealt{EGRETcalib})}, and improved observing efficiency (keeping the sky in the field of view with scanning observations, {\bf vs. inertial pointing for EGRET}). Pre-launch predictions of the instrument performance are described in \cite{LATpaper}.  Verification of the on-orbit response is in progress {\bf \citep{LATcalib}} but the indications are that 
it is  close to expectations.

The data analyzed for this source list were obtained during 4 August 2008 -
30 October 2008 (LAT runs 239503624 through 247081608, where the numbers refer
to the Mission Elapsed Time, {\bf or MET}, in seconds since 00:00 UTC on 1 January 2001). During this
time $Fermi$ was operated in sky scanning survey mode (viewing direction rocking
35$\degr$ north and south of the zenith on alternate orbits, except for a few
hours of special calibration observations during which the
rocking angle was much larger than nominal for survey mode or the
configuration of the LAT was different from normal for science operations. 
Time intervals when the rocking angle was larger than 47$\degr$ have been
excluded from the analysis, because the bright limb of the
Earth enters the field of view (see below).  In addition, two short time intervals associated with gamma-ray bursts (GRB)
that were detected in the LAT have been excluded. These intervals
correspond to GRB 080916C (MET 243216749--243217979,  \citep{GRB080916C})
and GRB 081024B (MET 246576157--246576187). The total
live time included is 7.53 Ms, corresponding to 82\% efficiency after
accounting for readout dead time and for observing time lost to
passages through the South Atlantic Anomaly  ($\sim$13\%).

The standard onboard filtering, event reconstruction, and classification were applied to the data \citep{LATpaper}, and for this analysis the `Diffuse' event class{\bf \footnote{See http://fermi.gsfc.nasa.gov/ssc/data/analysis/documentation/Cicerone/Cicerone\_Data/LAT\_DP.html.}} is used.  This is the class with the least residual contamination from charged-particle backgrounds.  The tradeoff for using this event class is primarily reduced effective area, especially below 500~MeV.  Test analyses were made with the looser `Source' class cuts and these were found to be less sensitive overall than Diffuse class for source detection and characterization.

The alignment of the $Fermi$ observatory viewing direction with the z axis of the LAT was found to be stable during survey-mode observation \citep{LATcalib}.  
The instrument response functions -- effective area, energy redistribution, and point-spread function (PSF) -- used in the likelihood analyses described below were derived from {\bf GEANT4}-based Monte Carlo simulations of the LAT using the event selections corresponding to the Diffuse event class.  The Monte Carlo simulations themselves were calibrated prior to launch using accelerator tests of flight-spare `towers' of the LAT \citep{LATpaper}.  Consistency checks with observations of bright sources in flight data are in progress \citep{LATcalib}.  Early indications are that the effective area below 100~MeV was overestimated {\bf by as much as 30\% owing to pile-up effects in the detectors}.  The source detection and spectral fitting analyses {\bf described below use only data $>$200~MeV.  The impact of the lower-than-predicted effective area below 200~MeV is limited.  The Diffuse event class already had relatively little effective area below 200~MeV, and so the impact on sensitivity for source detection is small.}  Analyses of flight data suggest that the PSF is somewhat broader than the calculated Diffuse class PSF at high energies; the primary effect for the current analysis is to decrease the localization capability somewhat.

For the bright source analysis a cut on zenith angle
was applied to the Diffuse class events to limit the contamination from
albedo $\gamma$-rays from interactions of cosmic rays with the upper
atmosphere of the Earth.  These interactions make the limb of the Earth
(zenith angle $\sim$113$^{\circ}$  at the 565 km, nearly-circular orbit of
{\it Fermi}) an intensely-bright $\gamma$-ray source
\citep{Thompson1981}.  The limb is very far off axis in survey mode
observations, but during a small fraction of the time range included in
this analysis the rocking angle reached angles as great as 47$^{\circ}$ (see
above) and so the limb was only $\sim$66$^{\circ}$ off axis.  Removing events
at zenith angles greater than 105$^{\circ}$ affects the exposure calculation
negligibly but  reduces the overall background rate.  After these cuts, the
data set contains $2.8 \times 10^6$ $\gamma$-rays with energies $>$100~MeV.

Figures 1 and 2 summarize the data set used for this analysis.  The {\bf intensity} map of Figure~1 shows the dramatic increase at low Galactic latitudes of the brightness of the $\gamma$-ray sky.  Figure 2 shows the corresponding exposure map for the representative energy 1~GeV.  The average exposure is $\sim$1 Ms and nonuniformities are relatively small (about 30\% difference between minimum and maximum), with the deficit around the south celestial pole due to loss of exposure during passages of $Fermi$ through the South Atlantic Anomaly \citep{LATpaper}.

\begin{figure}
\epsscale{.80}
\plotone{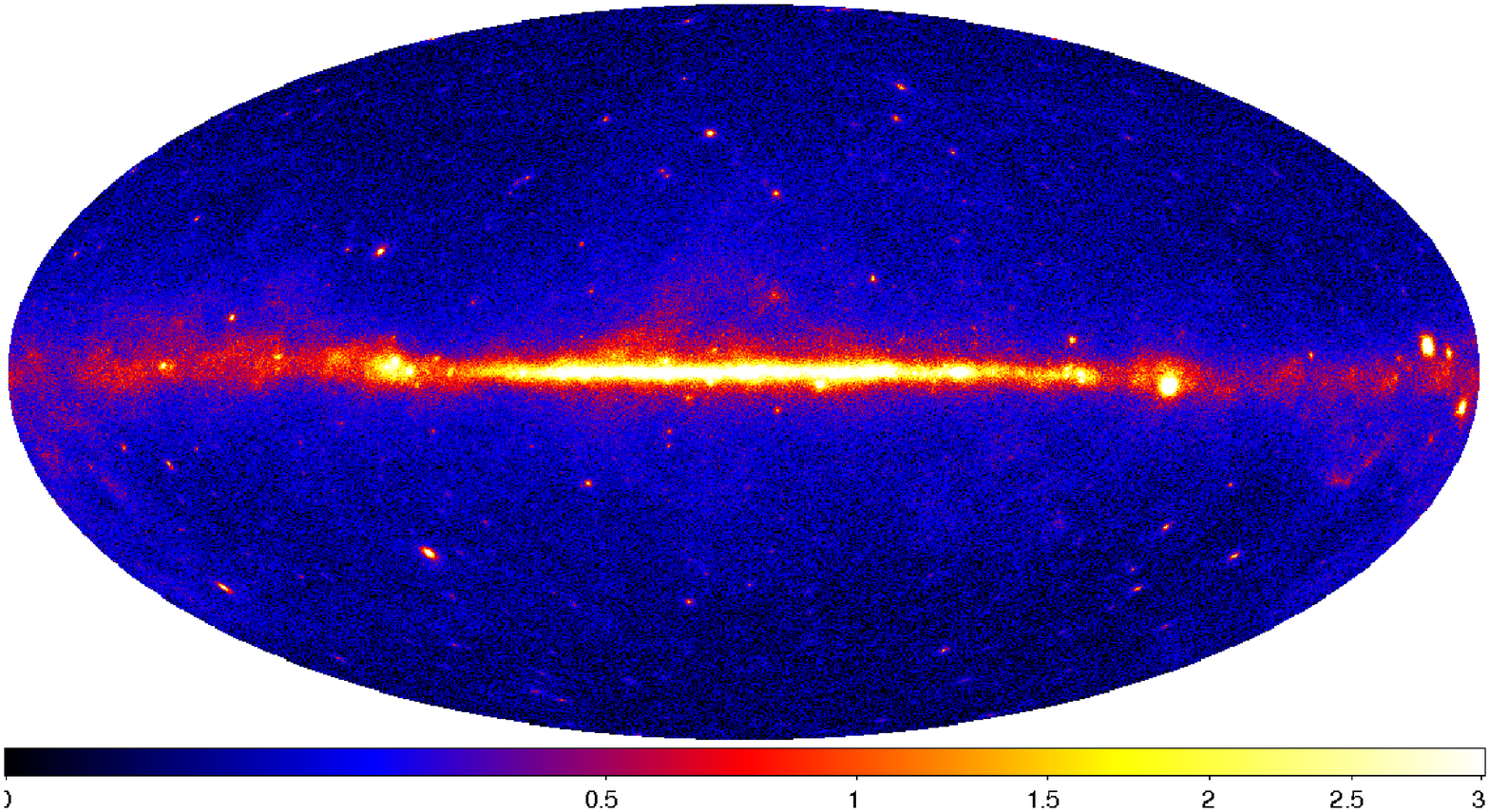}
\caption{{Sky map of the LAT data for the time range analyzed in this paper, Aitoff projection in Galactic coordinates. The image shows gamma-ray intensity for energies $>$300~MeV, in units of photons m$^{-2}$ s$^{-1}$ sr$^{-1}$.}\label{fig1}}
\end{figure}

\begin{figure}
\epsscale{.80}
\plotone{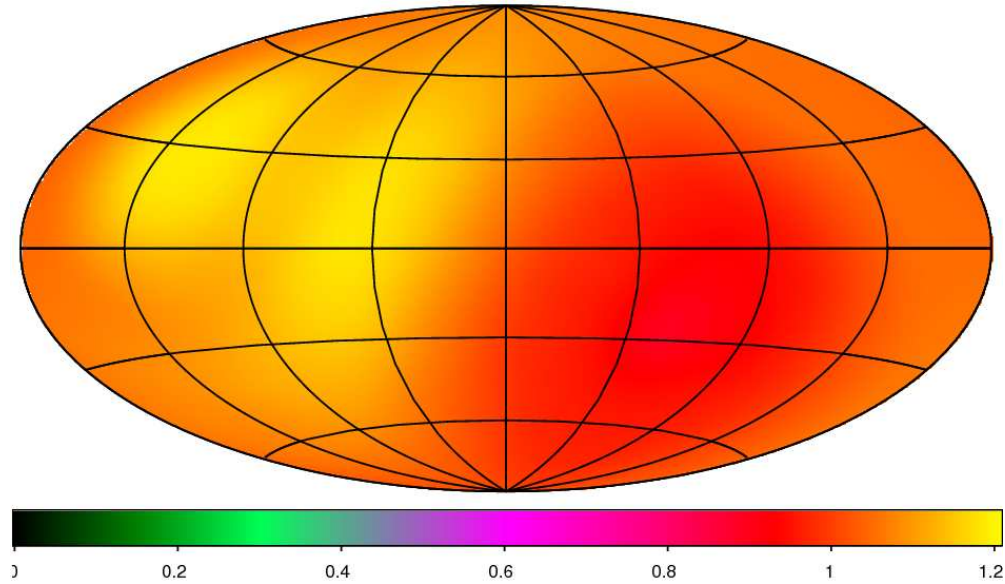}
\caption{{Exposure of the LAT for the time range analyzed in this paper, Aitoff projection in Galactic coordinates. The units are equivalent on-axis exposure in Ms. }\label{fig2}}
\end{figure}

\section{Construction of the Bright Source List}
\label{run:construction}

Although Figure~\ref{fig1} shows some obvious bright sources,
finding and measuring the properties of even the high-confidence sources
involves more than visual inspection of the map.
{\bf Because this analysis involves the entire sky and the broad energy range
of the LAT, it is necessarily more complex than the analysis
of an individual source.

The source list was built on the basis of the full
{\bf time interval}.
That is, we did not attempt to detect
{\bf potentially flaring}
sources on shorter time scales,
although we did check for variability of the sources
(Sect.~\ref{run:variability}) after the list was constructed.
Three steps were
applied in sequence}:
detection, localization, significance estimate.
At each step only a subset of the list at the previous step was kept.
In that scheme the bright source list threshold is defined at the last step,
but the completeness is controlled by the first one.
After the list was defined we determined the source characteristics
(flux in two energy bands, time variability) and we searched for possible
counterparts.

\subsection{Detection}
\label{run:detection}

At this time we do not have a good way to look for sources directly
in the 3D space of position and energy so we used standard image detection
techniques on counts images integrated over energy,
{\bf in which each event is simply stacked into the pixel
corresponding to its best-guess incident direction}.
The algorithm we used ({\it mr\_filter}) is based on wavelet analysis 
in the Poisson regime \citep{sp98}. It looks for local deviations from
the background model, leaving the background normalization free.
{\bf We used the same background model defined in Sect.~\ref{run:TS}, but
without any spectral correction.}
It returns a map of significant
features (above some threshold) on which we run a peak-finding algorithm,
SExtractor \citep{SExtractor}, to end up with a list of sources.
We also used for comparison another wavelet algorithm,
\citep[{\it PGWave,\rm}][]{dmm97,PGWave},
which differs in the detailed implementation and returns directly
a list of sources. Pre-launch simulations have shown that the latter was
somewhat more sensitive on a flat background
({\bf i.e.,} at high Galactic latitudes) but
did not work as well in the Galactic plane. At the 10-$\sigma$ level, the two
detection methods yield identical source lists.

An important decision was which energy
bands to use when applying the detection algorithms. The most important
instrumental characteristic in this respect is the point-spread function.
The 68\% containment radius improves by a factor of 25, 
from $\sim5\degr$ at 100~MeV to better than $1\degr$ at 1~GeV, reaching
$\sim0.2\degr$  above 10~GeV \citep{LATpaper}.
For this reason there is (at least over 3 months) little confusion
above 1~GeV and the diffuse background is not very limiting
except in the Galactic ridge. On the other hand,
most of the photons (83\%) are recorded below 1~GeV.
The majority of the sources
in the Galactic plane have overlapping PSFs and are background dominated
below 1~GeV (i.e., there
are more background than source events inside the PSF).
The starting energy therefore represents a trade between statistics
and resolution.

Another important aspect is that the events converted in the top, 
thin layers
of the tracker ($Front$ events) have nearly a factor of two better
PSF at a given
energy than those converted in the bottom thick layers
($Back$ events).
{\bf This corresponds to $Back$ events of energy $E$ having the same PSF width
as $Front$ events of energy $E/2$.}
Therefore, to optimize the sensitivity of the source detection we
{\bf used separate}
energy selections for $Front$ and $Back$ events.

{\bf The final scheme combines three energy bands.}
The full detection band ($1.8 \times 10^6$ events) starts at 200~MeV for $Front$
and 400~MeV for $Back$ events.
The remainder of the $2.8 \times 10^6$ events above 100~MeV carry
little position information and were not used for source detection.
We use a medium band starting at 1~GeV for $Front$ and 2~GeV for
$Back$ events ($3.2 \times 10^5$ events), which provides better position
estimates for hard spectrum sources.
We have also used a high energy band starting at 5~GeV for $Front$
and 10~GeV for $Back$ events. This band is very photon starved
($3 \times 10^4$ events)
but has essentially no background in a PSF-sized region and can be useful
for very hard sources and to avoid confusion in the Galactic plane.
We use smaller image pixels at high energy (0.1$\degr$) than in the medium
band (0.2$\degr$) and the full energy band (0.3$\degr$)
to adapt to the broader PSF at low energy.
The bands are not exclusive (i.e., the full band includes the high
energy photons) because the high-energy events always improve the detection.
To obtain a global list of candidate sources we start with the 
sources detected in the high-energy band (best localization)
and add the sources detected in the lower-energy bands in turn,
excluding sources whose positions are consistent with detections
at higher energies.

{\bf Because the source detection methods are standard algorithms not
specific to {\it Fermi} they} work in cartesian coordinates, not the
spherical sky. We map the whole sky with 24 local World Coordinate System
projections \citep{cg02}
in Galactic coordinates: 4 CAR (plate carr\'ee) projections
along the Galactic plane
covering $-$10$\degr$ to +10$\degr$, 6 AIT(Hammer-Aitoff) projections
on each side of the plane
covering 10$\degr$ to $45\degr$, and 4 ARC (zenithal equidistant) projections
(rotated $45\degr$ so that
the pole is in a corner) covering 45$\degr$ to 90$\degr$. Each map is $5\degr$
larger on each side than the area from which the sources are extracted,
to avoid border effects.

We set the threshold of the source detection step at 4-$\sigma$.
This resulted in 562 `seed' sources. 290 were best detected in the full band,
212 in the medium band and 60 in the high band (among 151 total excesses above
4-$\sigma$ in that band).

\subsection{Localization}
\label{run:localization}

The image-based detection algorithms provide estimates of the source
positions, but the positions are not optimal because
{\bf the energy-dependent extent of the PSF is not fully taken into account}.
These methods also do not supply error estimates on the positions.

The method that we use to localize the sources ({\it pointfit}) is a
{\bf binned}
likelihood technique. It uses relatively narrow 
energy bins (typically 4 per decade) and sums log(likelihood) over the energy bins. It does not
use events below 500~MeV, which carry little information on position.
To optimize the technique further the analysis gathers $Front$ and
$Back$ events
according to their PSF widths rather than their energies.
Each source is treated independently. That means the model
is a point source (with the same position but different width in each
energy bin following the PSF) on top of a background model with free
scaling in each
{\bf energy}
bin.
{\bf The sources are treated in descending order}
such that
brighter sources are included in the background model for fainter
{\bf ones. The closest nearby source was 0.5$\degr$, with only a small effect
on the fits to the lower energy bins.
The program} 
returns the best-fit position and the error estimate (1-$\sigma$
along one dimension)
{\bf based}
on the assumption that $-2\Delta$log(likelihood) behaves
as a $\chi^2$-distribution.
{\bf The LAT PSF itself is very close to axisymmetric \citep{LATpaper}.
The error box is not in general circular due to fluctuations in the positions
of the few high energy photons that dominate
the localization precision. Here we neglect this effect, which is small for
strong sources, and provide only error circles.}

Of the 562 initial sources, {\it pointfit} did not converge for 50
at this step, or converged to another nearby source.
{\bf The reason could be confusion, or a soft spectrum leading to
too few source events above 500~MeV.}
We did not discard those outright, but
kept their original positions. Several of them were deemed significant
by the maximum likelihood algorithm (Sect.~\ref{run:TS}). We defined
the positions and position uncertainties of those using a more precise
but much slower tool ({\it gtfindsrc}) which accounts for all sources
in the vicinity.
{\bf More precisely, we included in the local model all nearby sources
(even those below the bright source limit defined in Sect.~\ref{run:TS}).
The spectral parameters of those within 1$\degr$
of the current source were left free, but
only the current source's position was adjusted in a given run.}
The same tool was used in a number of confused
regions (mostly close to the Galactic plane)
in which the primary analysis did not converge well.
32 sources in all were treated that way,
including 13 of the bright sources presented here.
In the end 532 sources survived the localization step.

The angular uncertainties for localization are determined from the shape
of the likelihood function as described above.
{\bf This results in a 1-dimensional 1-$\sigma$ error estimate $\Delta x_{\rm stat}$.
For a 2D axisymmetric gaussian distribution the 95\% confidence level radius
$r_{95}$ is related to $\Delta x_{\rm stat}$ by a factor 
$\sqrt{-2\log (1-0.95)}=2.45$.
However examining the distribution of the position errors from high-confidence,
identified sources,
we found that we needed to increase the uncertainties by 40\%
in order to be sure of including 95\% of the cases.}

For very bright sources like Vela, the observed offsets from the true position
observed with the present analysis led us to add in quadrature
an additional systematic uncertainty of 0.04 degrees to $r_{95}$.
\begin{equation}
\label{eq:sysposerr}
r_{95}^2 = (1.4 \times 2.45 \times \Delta x_{\rm stat})^2 + (0.04\degr)^2
\end{equation}
Both the 1.4 correction factor and the $0.04\degr$ systematic uncertainty
are conservative and are expected to improve.

\begin{figure}
\label{fig3}
\epsscale{.80}
\plotone{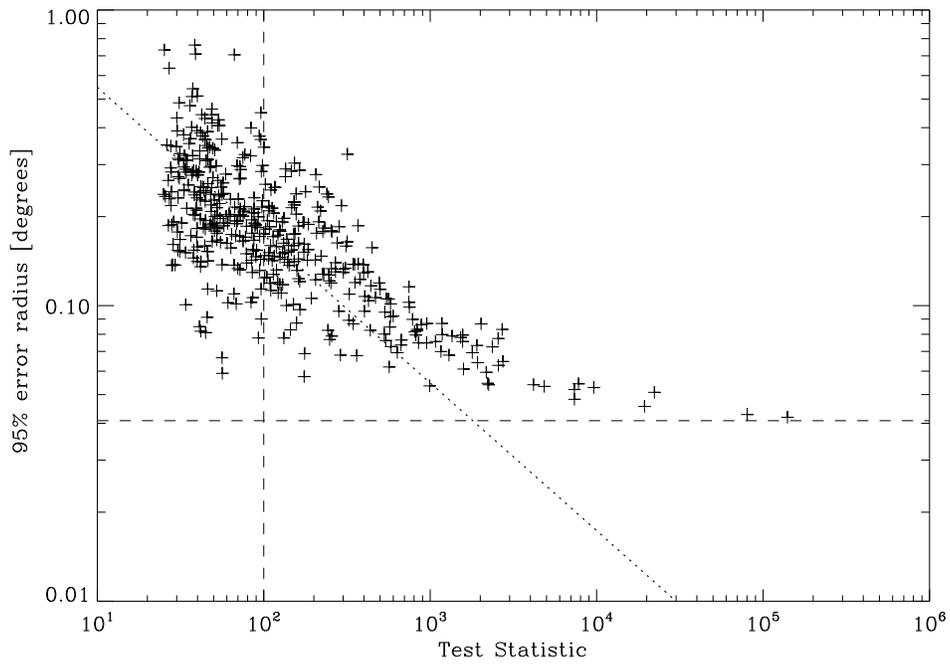}
\caption{Source location uncertainty radii ($r_{95}$ from Eq.~\ref{eq:sysposerr})
as a function of Test Statistic (Sect.~\ref{run:TS}),
down to a limit of $TS = 25$.
The dotted line is a 1/$\sqrt{TS}$ trend for reference.
The vertical dashed line is our $TS$ = 100 threshold.
The horizontal dashed line is the
{\bf absolute}
systematic error that we adopted.}
\end{figure}

Figure 3 illustrates the resulting position uncertainties 
as a function of the Test Statistic $(TS)$ values obtained in
Sect.~\ref{run:TS}.
The relatively large dispersion that is seen at a given $TS$
is in part due to the different local conditions
(level of diffuse $\gamma$-ray emission)
but primarily to the source spectrum. Hard sources are better localized
than soft ones for the same $TS$ because the PSF is so much
arrower at high energy.
At our threshold of $TS$ = 100 (10-$\sigma$) the typical 95\% uncertainty radius is
about $10\arcmin$, and the maximum is $20\arcmin$.

\subsection{Significance and Thresholding}
\label{run:TS}

The detection and localization steps provide estimates of significance,
but these are underestimates because the detection step does not
explicitly use the energy information and the localization step does not use
the low energy events.
To better estimate the source significances we use a 3D
maximum likelihood algorithm ({\it gtlike}) in unbinned mode
{\bf i.e.,}
each event is considered individually
{\bf according to its direction, energy, and conversion location in the LAT.} 
This is part of the standard Science Tools software package\footnote{http://fermi.gsfc.nasa.gov/ssc/data/analysis/documentation/Cicerone/}
currently at version 9r9.  The {\it gtlike} tool provides for each source
the best fit parameters
and the Test Statistic
$TS=2\Delta$log(likelihood) between models
with and without the source.
That tool does not vary the
source position, but it adjusts the source spectrum.  It should be noted
that
{\bf \it gtlike}
does not include the energy dispersion in the
$TS$ calculation
{\bf (i.e., it assumes that the measured energy is the true energy)}.
Given the 8\% to 10\% 1-$\sigma$ energy resolution of the LAT over the energy bands
used in the present analyses, this approximation is justified.  The underlying
optimization engine is Minuit\footnote{http://lcgapp.cern.ch/project/cls/work-packages/mathlibs/minuit/doc/doc.html}.
The code works well with up to $\sim$30 free parameters,
an important consideration for regions where sources are close enough
together to partially overlap.
Uncertainty estimates (and a full covariance matrix) are obtained from Minuit
in the quadratic approximation around the best fit.
For this stage we modeled the sources with simple power-law spectra.

The $TS$ associated with each source is a measure of the source significance,
or equivalently the probability that such an excess can be obtained
from background fluctuations alone.
The probability distribution in such a situation (source over background)
is not precisely known \citep{pdc02}.
However since we consider only positive fluctuations, and each fit
involves two degrees of freedom (flux and spectral slope), 
the probability to get at least $TS$ at a given position in the sky
is close to 1/2 of the $\chi^2$ distribution with two degrees of freedom
\citep{EGRET_like}, so that $TS$ = 25
corresponds to 4.6 $\sigma$ (one sided).
Pre-launch simulations have shown that this {\bf approximation} is indeed true if the
background model is close to the truth.

The diffuse background is of course very important since it represents
around 90\% of the events. We model the Galactic diffuse emission
using GALPROP, described in \citet{SMR} and \citet{Strong07},
which uses a realistic representation of cosmic-ray propagation
in the Galaxy and the resulting $\gamma$-ray emission;
it uses distributions of gas based on radioastronomical surveys,
and the interstellar radiation field from an extensive modelling package.
For this work, the GALPROP package has been updated to include recent
H~I and CO surveys, more accurate decomposition into Galactocentric rings,
{\bf as well as a new calculation of the interstellar radiation field
for inverse Compton emission \citep{PorterISRF}.
For this 
work the fit of the model to the {\it Fermi} data was improved by an increase in
the inverse Compton component, and a flatter cosmic-ray gradient in 
the outer Galaxy.}
The particular GALPROP run designation
for our model is \texttt{54$_-$59varh7S}.

Because the fitted fluxes and spectra of the sources
can be very sensitive to even slight errors in the spectral shape of the
diffuse emission we allow
{\bf the Galactic diffuse model}
to be corrected (i.e., multiplied) locally
by a power law in energy with free normalization and 
{\bf spectral}
slope.
The slope varies between 0 and 0.15 (making it harder) in the Galactic plane
and the normalization by $\pm$ 20\%.
The isotropic component of the diffuse emission
{\bf represents the}
extragalactic and residual backgrounds
{\bf (instrumental + Earth albedo). It}
is modeled by a simple power law. Its 
{\bf spectral}
slope was fixed to $E^{-2.25}$,
the best fit value at high latitude, and its normalization was left free.
{\bf The three free parameters were separately adjusted in each
Region of Interest (see below).}

For this significance analysis,
we used only events with energies above 200~MeV,
because the fits to the diffuse
spectrum were systematically high below 200~MeV; the extrapolation
of the high-energy spectrum overestimated the data,
{\bf possibly because of the acceptance bias described
in Sect.~\ref{run:systematics}}.
We feared that
including the low-energy points could bias the whole process.
This energy cut changes little the $TS$ estimates
except for the very softest sources.
The high energy limit for the analysis was set to 100~GeV.
{\bf There were fewer than 1000 events above 100~GeV, and at this point
we do not have a single source that is bright enough to check
our calibration above that limit.}

We split the sky into overlapping circular Regions of Interest (RoI),
each typically $15\degr$ in radius. The source parameters are free
in the central part of each RoI
({\bf which is chosen}
such that all free sources are well within
the RoI even at low energy). We adjust the RoI size so that not more than
8 sources are free at a time. Adding 3 parameters for the diffuse
model, the total number of free parameters in each RoI is 19 at most.
We needed 128 RoIs to cover the 532 seed positions.

We proceed iteratively.
{\bf All RoIs are processed in parallel and a global current model
is assembled after each step in which the best fit parameters
for each source are taken from the RoI whose center is closest to the source.
At each step the}
parameters of the sources close to the borders are fixed to their
{\bf values in the global model at the end of}
the previous step; they
{\bf all}
start at 0
{\bf flux at the first step (the starting point for the spectral slope is 2)}.
Sources formally outside the RoI (but which can contribute at low energy
due to the broad PSF) are included in the model as well.
We iterate over 5 steps (the fits change very little after the fourth).
At each step we remove sources with low $TS$
{\bf and refit,}
raising the threshold
up to 25 (approximately 5-$\sigma$) at the last step.
{\bf We have checked via simulations that removing the faint sources
has little impact on the bright sources, much less so than changing
the diffuse model does (Sect.~\ref{run:systematics}).}

This procedure left 444 sources, among which 205 have $TS >100$.
We chose not to include the lower-significance sources ($TS < 100$)
in the bright source list for two reasons:
\begin{itemize}
\item The number of sources per Test Statistic interval normally decreases
with 
{\bf increasing}
$TS$ for any logN-logS close to Euclidean. This is not the case
with our procedure (there are fewer sources at $25 < TS < 30$
than at $35 < TS < 40$)
particularly in the Galactic plane. This is a rather sure sign
that we are missing sources at low flux, and more so in the Galactic plane.
Given the relatively rough nature of the detection procedure
(Sect.~\ref{run:detection}) this is not particularly surprising.
\item Judging by the spatial and spectral residuals the Galactic diffuse
model is still in need of improvement (see Sect.~\ref{run:systematics}).
This uncertainty makes us wary of claiming detections
of sources not too far above the diffuse level.
\end{itemize}

On the other hand the sources at $TS > 100$ can be seen by eye
on the images and we are confident they are all real.
Note that all excesses formally above $TS = 25$
were included in the maximum likelihood adjustments
(including those described in Sect.~\ref{run:flux} and ~\ref{run:variability})
to avoid transferring their fluxes to the more-significant sources.

Figure~4 shows the source flux needed to reach a 10-$\sigma$ 
significance level at any point in the sky for the 3-month time interval 
considered in this analysis. 
This is based on a calculation using the Galactic diffuse and 
isotropic background models, the instrument response functions of the LAT,
and the pointing history during the 3 months, and assuming an $E^{-2.2}$
spectrum, the average spectral shape of the sources.
This should be viewed as an indication
{\bf only}
because the detection threshold
depends on the source spectrum.
Although the nonuniform exposure affects this map somewhat,
the dominant factor is the strong diffuse emission along the Galactic plane.

\begin{figure}
\label{fig4}
\epsscale{.80}
\plotone{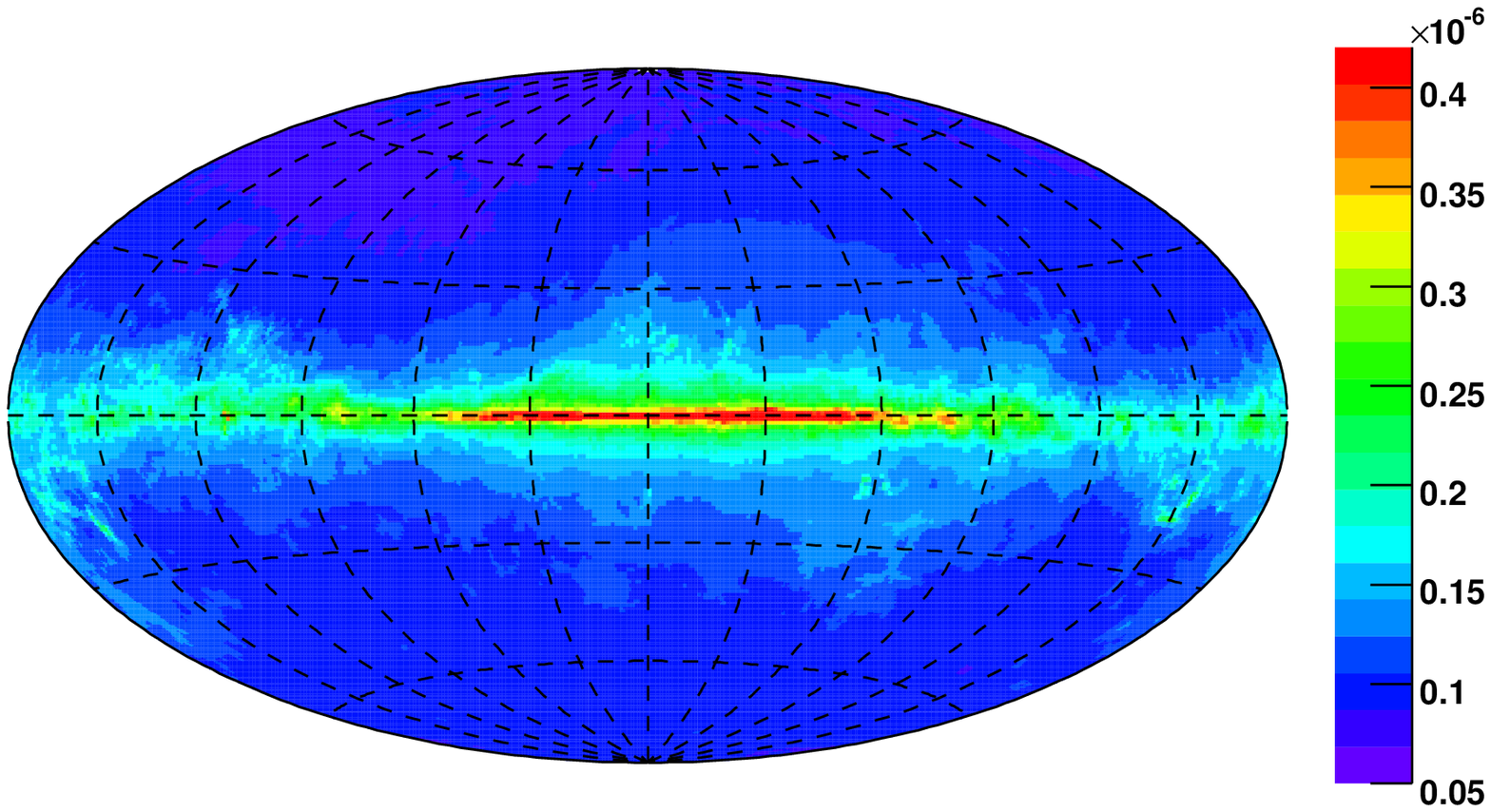}
\caption{Flux (E$>$100~MeV
{\bf in ph cm$^{-2}$ s$^{-1}$})
needed for a 10-$\sigma$ detection for the LAT data for the 3-month time range
considered in this paper. The assumed photon spectral index is 2.2.
Galactic coordinates.}
\end{figure}

\subsection{Flux Determination}
\label{run:flux}

The maximum likelihood method described in Sect.~\ref{run:TS}
provides good estimates of the source significances, but not very
accurate estimates of the fluxes. This is because the spectra of most sources
do not follow a single power law over that broad an energy range (more than
two decades). Among the two most populous
{\bf classes,}
the AGN often show a broken power-law spectrum and the pulsars
an exponentially cut off
{\bf power law.}
In both cases fitting a single power law over the entire range overshoots
at low energy where most of the photons are, and therefore biases
the fluxes high
{\bf (on the other hand the effect on the significance is low
due to the broad PSF and high background at low energies)}.
An additional difficulty is that the fit over the entire range stopped
at 200~MeV, whereas comparison with previous missions requires that
we provide fluxes starting at 100~MeV. Extrapolating back to 100~MeV
would have added another error.

To provide better estimates of the source fluxes, we have decided
to split the range in two and define two independent bands
from 100~MeV to 1~GeV and 1~GeV to 100~GeV.
{\bf The 1~GeV limit is largely arbitrary but is a round number
that happens to split the data into approximately equal
contributions to the sources' significance.}
The list of sources remains the same in the two bands of course.
Each band is treated in the same way as the full band in Sect.~\ref{run:TS}.
The power-law slopes are fitted independently
in each band for each source. We discard the slopes here because they are
not very precise (the low band is not very broad and there are not many
events in the high band) and keep only the flux estimates.
Even though the fit is not good near 100~MeV as mentioned in Sect.~\ref{run:TS}
(see also Sect.~\ref{run:systematics}), including the data down to 100~MeV
still provides a more reliable estimate of the flux than extrapolation
for all sources which do not follow exactly a power law.
The estimate from the sum of the two bands is on average within 30\%
of the flux obtained in the previous section, with excursions
up to a factor 2.
{\bf We have also compared those estimates with a more precise
spectral model for the three bright pulsars
(Vela, Geminga and the Crab). The flux estimates are within 5\% of each other.}

An additional difficulty that does not exist when considering the full data
is that, because we wish to provide the fluxes in both bands for all sources,
we must handle the case of sources that are not significant in one
of the bands or where the flux is poorly determined due to large uncertainty
in the spectrum.
This situation occurs even for the high-confidence sources reported here:
9 have $TS < 25$ for the 100~MeV - 1~GeV band, and 2 have $TS < 10$
{\bf in this band}.
No high-confidence source has $TS < 25$ in the  1-100~GeV band.
This difference reflects the fact that the current study
(and the LAT in general)
is more sensitive at high energy.
For the sources with $TS < 10$ or poorly-measured flux values
(where the nominal uncertainty is comparable to the flux itself),
we replace the flux value from the likelihood analysis
by a 2-$\sigma$ upper limit (2$\Delta$log(likelihood) = 4),
indicating the upper limit by a 0 in the flux uncertainty column
{\bf of Table \ref{table1}}.

\subsection{Variability}
\label{run:variability}

For this paper we wanted to flag sources that are clearly variable.
To that end we use the same energy range as in Sect.~\ref{run:TS}
(200~MeV to 100~GeV) to study variability.
To avoid ending up with too large error bars
in relatively short time intervals, we froze the spectral
index of each source to the best fit over the full interval.
Sources do vary in spectral shape as well as in flux, of course, but
we do not aim at characterizing source variability here, just detecting it.
It is very unlikely that a true variability in shape will be such
that it will not show up in flux at all.

We split the full three-month interval into $N_{\rm int}$ = 12 intervals
of a little more
than one week. This preserves some statistical precision for the
moderately bright sources we are dealing with here, while being sensitive
in the right time
{\bf scale}
for flaring blazars.
Because we do not expect the diffuse emission to vary, we freeze
the spectral adjustment of the Galactic diffuse component to the local
(in the same RoI)
best fit over the full interval.
We need to leave
{\bf the normalization of at least one diffuse component}
free (just to 
adapt to the natural Poisson variations of the background).
Because it was not obvious which one to freeze we decided to leave both
(Galactic and isotropic) free in each interval.
So in the end the fitting procedure is the same as in Sect.~\ref{run:TS}
except that all spectral shape parameters are frozen.
The faint sources were left free
{\bf (even when not significant in the current interval)}
as well as the bright ones.

As in Sect.~\ref{run:flux} it often happens that a source is not
significant in all intervals. To preserve the variability index
(Eq.~\ref{eq:varindex}) we keep the best fit value and its estimated
error even when the source is not significant. This does not work,
however, when the best fit is very close to zero
{\bf because in that case the log(likelihood) as a function of flux
is very asymmetric}.
Whenever $TS < 1$
we compute the 1-$\sigma$ upper limit and replace the
error estimate with
{\bf the difference between that upper limit and the best fit.
This is an estimate of the error on the positive side only.
The best fit itself is retained}.

\begin{figure}
\label{fig5}
\epsscale{.80}
\plotone{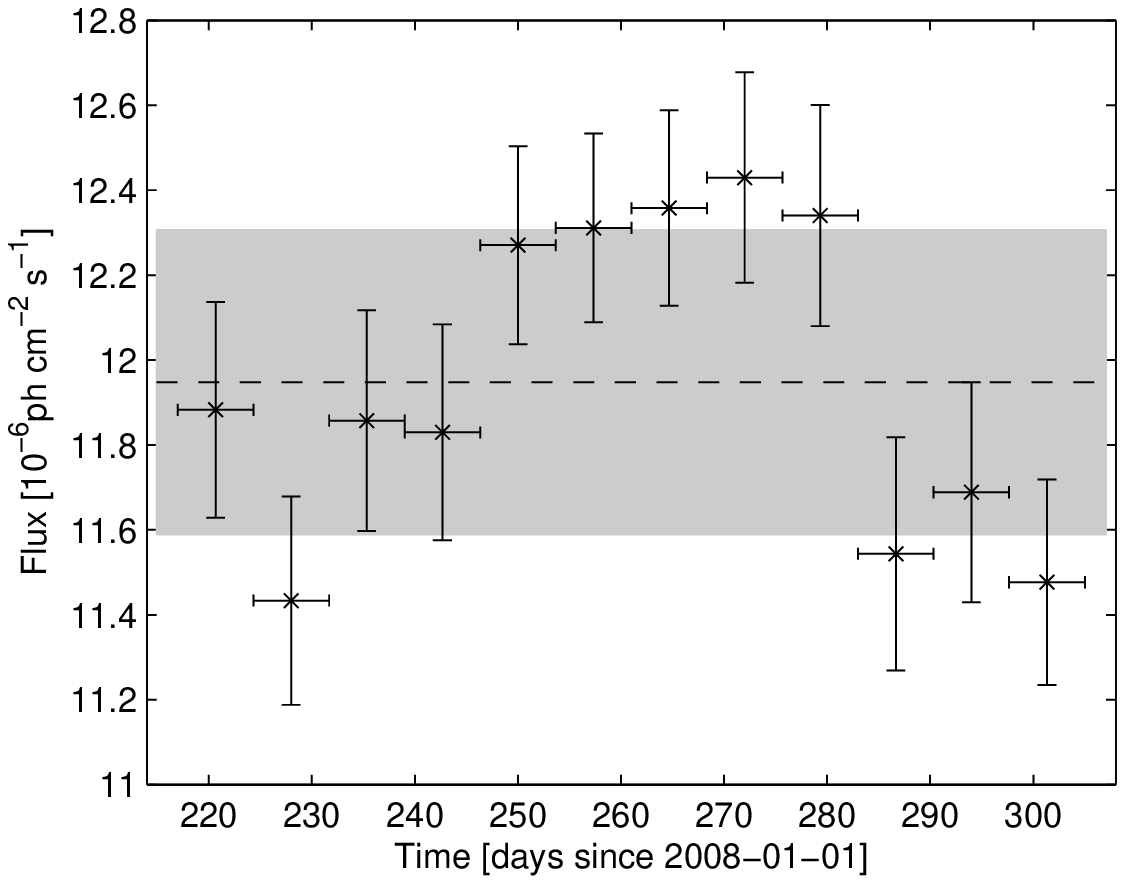}
\caption{Light curve of Vela (0FGL J0835.4$-$4510)
with fluxes from a single power-law fit and purely statistical error bars.
Each interval is approximately one week.
The dashed line is the average value.
Because Vela is very bright it would
{\bf have been classified as variable using the statistical errors only},
but the flux dispersion is only
{\bf 2.3\% beyond statistical}.
The grey area shows the 3\% systematic error we have adopted.
Note that because this analysis uses a power-law model over the entire range
from 200~MeV to 100~GeV
it grossly overestimates the true flux of Vela, but this effect
does not depend on time.}
\end{figure}

\begin{figure}
\label{fig6}
\epsscale{.80}
\plotone{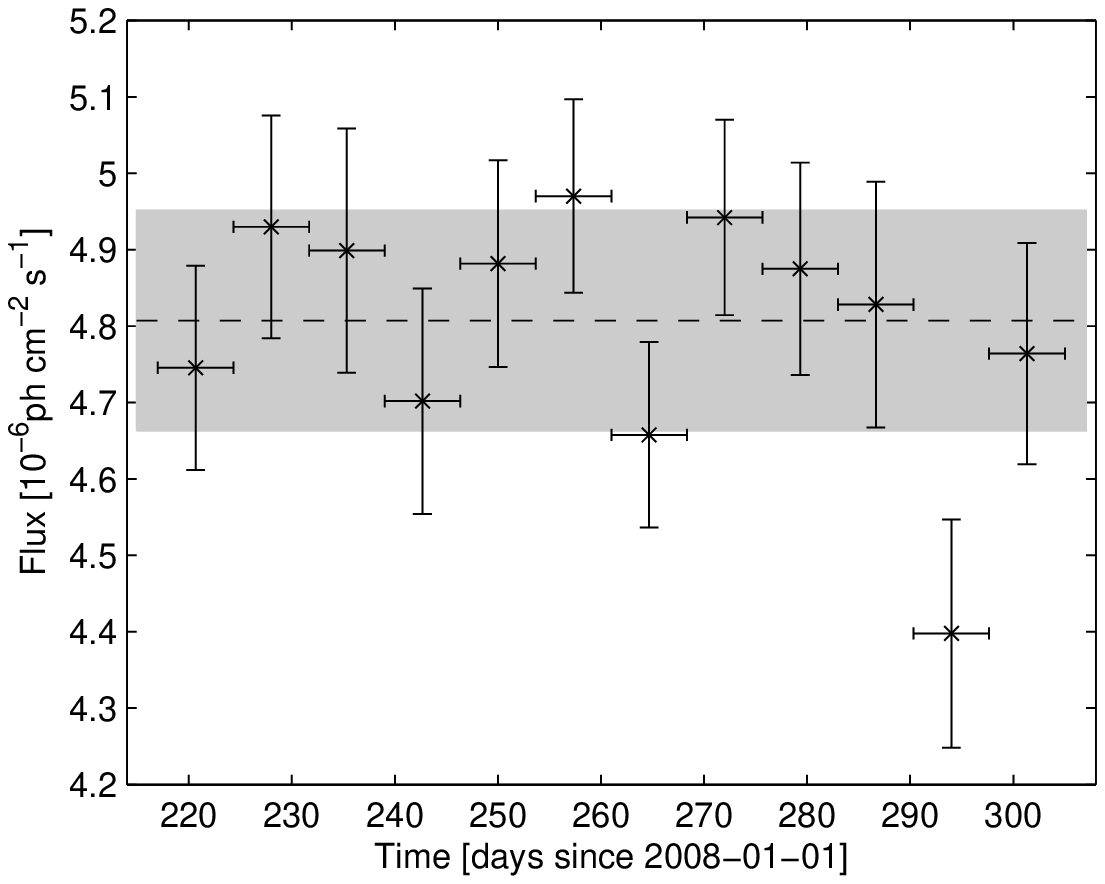}
\caption{Same as Figure 5 for Geminga (0FGL J0634.0+1745).
{\bf That pulsar's flux dispersion is 1.6\% beyond statistical.
Its variability index (Eq.~\ref{eq:varindex}) would
not have exceeded the threshold even with pure statistical errors.}}
\end{figure}

Figures 5 and 6 show the fluxes derived for the Vela and Geminga pulsars
as a function of time. 
As the brightest persistent sources, Vela and Geminga provide a reference
for non-variability.
Based on these light curves, we estimate that the instrument
and processing (event classification) are stable on time scales of weeks
to 2\% relative precision. 
To be conservative we have added in quadrature
a fraction $f_{\rm rel}$ = 3\% of the average flux $F_{\rm av}$
to the error estimates
{\bf (for each 1-week time interval)}
used to compute the variability index. 

Figure 7 shows the flux derived for the AO 0235+164 blazar
as a function of time.  In contrast to the steady pulsars,
many of the blazars detected by the LAT show strong variability. 

\begin{figure}
\label{fig7}
\epsscale{.80}
\plotone{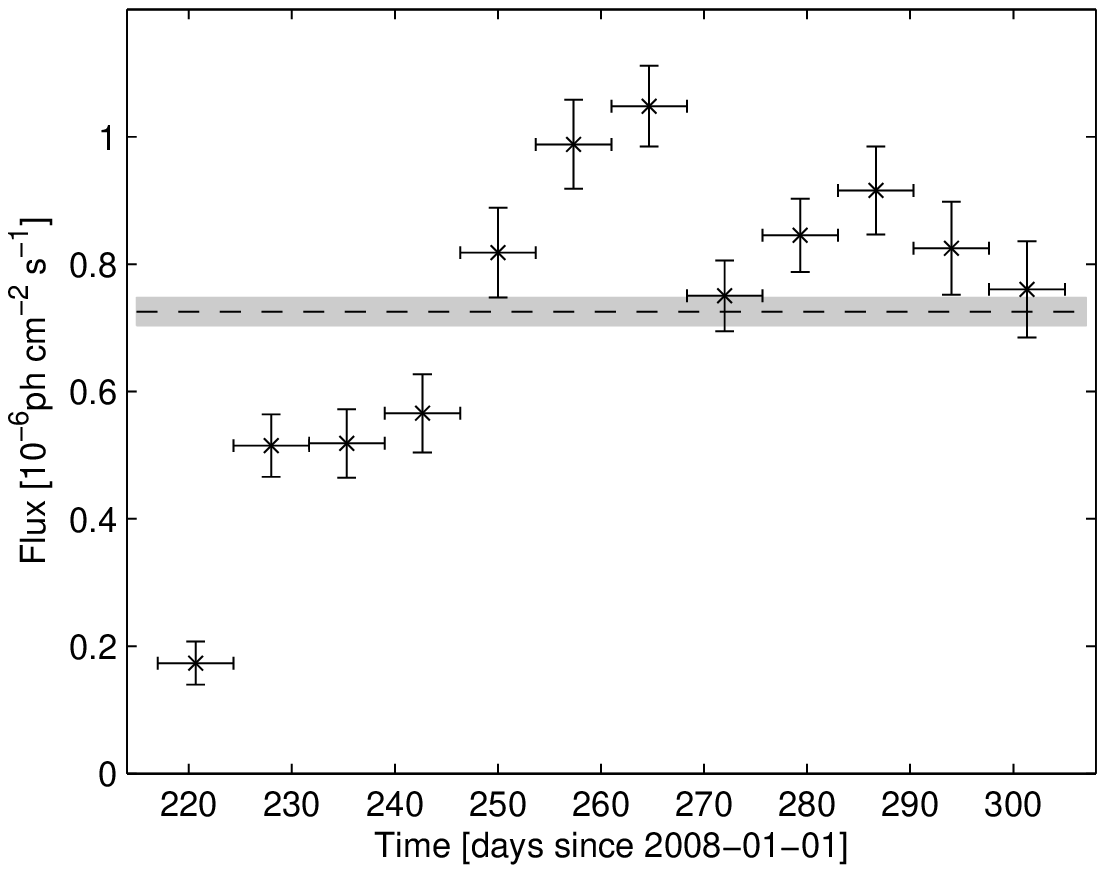}
\caption{Same as Figure 5 for AO 0235+164 (0FGL J0238.6+1636),
a variable blazar.
Note the difference in scale from the Vela and Geminga light curves.}
\end{figure}

The variability index is defined as a simple $\chi^2$ criterion:
\begin{equation}
\label{eq:varindex}
V = \sum_i \frac{(F_i - F_{\rm av})^2}{\sigma_i^2 + (f_{\rm rel} F_{\rm av})^2}
\end{equation}
where $i$ runs over the 12 intervals and
$\sigma_i$ is the statistical uncertainty in $F_i$.
Since $F_{\rm av}$ is not known a priori, this parameter is expected,
in the absence of variability, to follow a $\chi^2$ distribution
with 11 (= $N_{\rm int} -$ 1) degrees of freedom.
We set the variability flag True whenever the probability of getting
the value of $V$ or more by chance is less than 1\% (so that we expect
2 false positives over the sample of 205 sources).
This corresponds to $V > 24.7$.  This variability index is robust
for the bright sources considered here,
although for less-significant sources methods that handle upper limits
will be needed.

\begin{figure}
\label{fig8}
\epsscale{.80}
\plotone{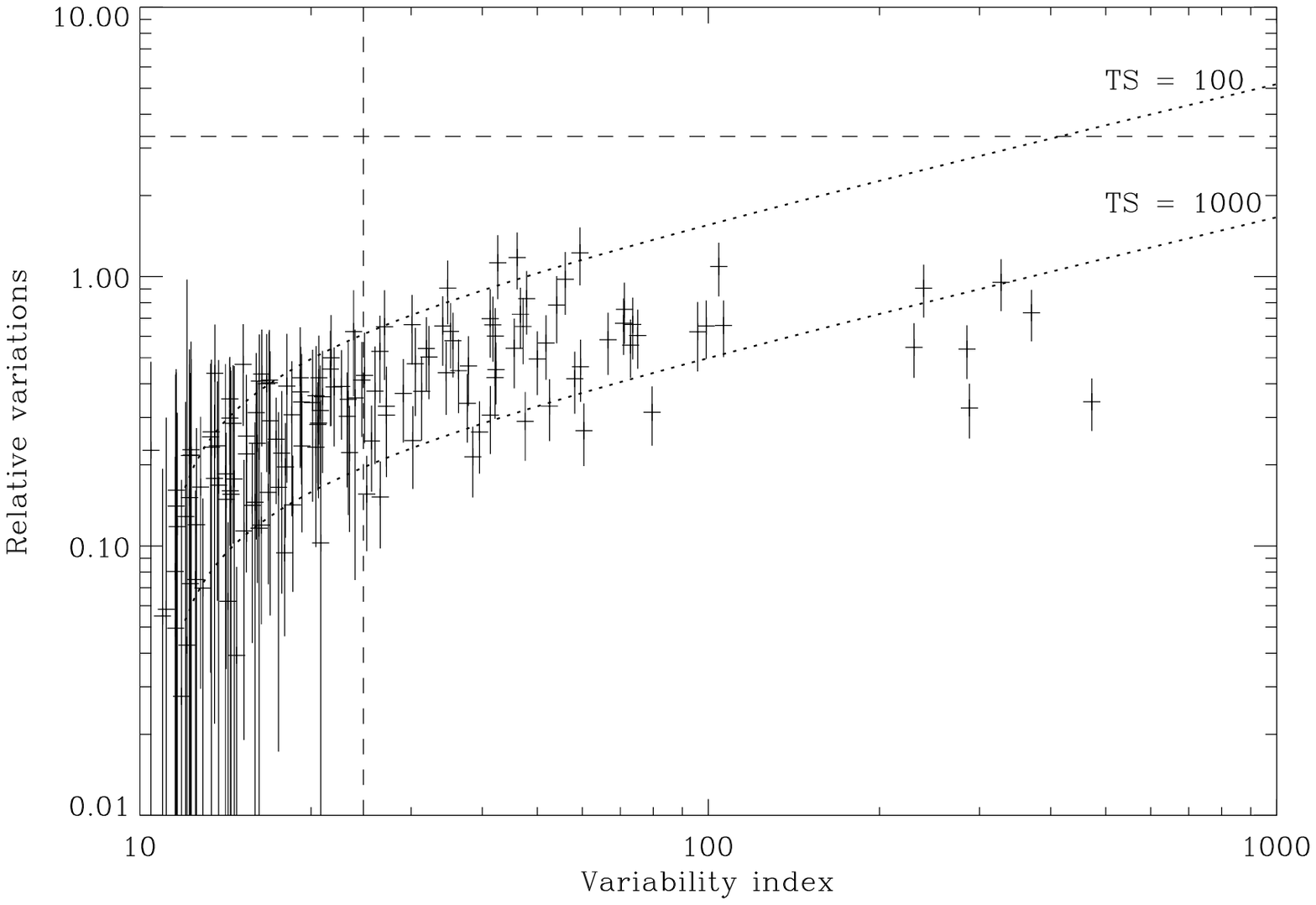}
\caption{Relative source variability plotted as a function
of the variability index (Eq.~\ref{eq:relvar}).
The vertical dashed line shows where we set
the variable source limit. The horizontal dashed line
is the maximum relative variability that can be measured
$\sqrt{N_{\rm int}-1}$.
The dotted lines show how the variability index depends on $\delta F/F$
at our threshold ($TS$ = 100) and for brighter sources ($TS$ = 1000).
At a given $TS$, the lower right part of the diagram is not accessible.
{\bf For more details see Sect.~\ref{run:variability}.}}
\end{figure}

Figure 8 shows the relative variability of the sources.
It is defined from the excess variance on top of the statistical
and systematic fluctuations:
\begin{equation}
\label{eq:relvar}
\delta F/F = \sqrt{\frac{\sum_i (F_i - F_{\rm av})^2}
                        {(N_{\rm int}-1) F_{\rm av}^2} -
                   \frac{\sum_i \sigma_i^2}{N_{\rm int} F_{\rm av}^2} -
                   f_{\rm rel}^2}
\end{equation}
The typical relative variability is 50\%, with only a few strongly variable
sources beyond $\delta F/F$ = 1.
The dotted lines show how the relative variability depends on the
variability index as a function of $TS$, assuming that the relative precision
on flux is the same for all sources (we get
{\bf a relative precision of 0.16 for the flux}
on average at $TS$=100).
{\bf So this means that the}
criterion we use is not sensitive to relative variations smaller
than
{\bf 60\%}
at $TS$ = 100.
{\bf That limit goes down to 20\% as $TS$ increases to 1000.}
Note that because the relative precision on flux is not exactly the same
for all sources the cutoff,
{\bf expressed in relative variability,}
is not sharp.
It is clear that we must be missing many variable AGN below $TS$ = 1000.

66 bright sources (one third of the sample) are declared variable.
This is far more than the 2 false positives expected,
so most of them are truly variable.
This level of variability is not particularly surprising,
as blazars are known to be
strongly variable on timescales of days to weeks.
We emphasize that sources not flagged may also show variability
at lower amplitude or different time scales than used for this test.
We refer to these other sources as ``non-variable'' (on weekly time scales) rather than ``steady.''

\subsection{Limitations and Systematic Uncertainties}
\label{run:systematics}

A limitation of this work is that we
did not attempt to test for source extension. All sources are
assumed to be point-like. This is true for all known source populations
in the GeV range (see Sect.~\ref{run:assoc}).
On the other hand the TeV instruments
have detected many extended sources in the Galactic plane, mostly pulsar
wind nebulae  and supernova remnants, 
\citep[e.g.,][]{HESS_Plane,Milagro}. 
The current level of LAT exposure cannot address source extension
at the level seen by the TeV telescopes. 

We have addressed the issue of systematics for localization
in Sect.~\ref{run:localization}. This section deals with the systematic
uncertainties on flux estimates.
{\bf An obvious one is the power-law representation within each energy band.
If one source had a very curved spectrum (like a spectral line) its
flux estimate certainly would be inaccurate. Our experience with those
sources for which more detailed studies have been made, though, is that
the current estimates are fully acceptable.
Beyond that, there}
are two main sources of systematics:
the imperfect knowledge of the instrument so early into the mission,
and the imperfect modeling of the diffuse emission.

The fluxes are calculated using pre-launch calibrations (designated P6\_V1) based on
Monte Carlo simulations and a beam test at CERN \citep{LATpaper}.
In flight, the presence of pile-up signals in the LAT tracker and calorimeter
left by earlier particles was revealed in periodic trigger events. This effect
leads to a reduction of the actual acceptance as compared to the pre-launch
prediction, as fewer events pass the rejection cuts, most notably for  photons
below 300~MeV. The magnitude of this reduction is still under investigation,
but the fluxes reported here may be lower than the true ones by as much as
35\% below 1~GeV and 15\% above 1~GeV. Because of the current uncertainty, no
correction has been applied to the results; these effects are being assessed
in detail, and will then be included in a reprocessing of the data. This
uncertainty applies uniformly to all sources.
Our relative errors (comparing one source to another
or the same source as a function of time) are much smaller,
as indicated in Sect.~\ref{run:variability}.

It is interesting to note that the flux above 100~MeV that the LAT finds
for the three historical pulsars (Vela, Geminga and the Crab) is actually
very close to that reported in the 3EG catalog \citep{3rdCat}.
Geminga and the Crab are within 1 $\sigma$, and the LAT flux for Vela
(9.15 $\times 10^{-6}$ ph cm$^{-2}$ s$^{-1}$) is 11\% higher
than that of EGRET. This implies that the bias may be on the low
side of our estimate, unless EGRET also underestimated the source flux.

The diffuse emission is the other important source
of uncertainties. Contrary to the former, it does not affect all sources
equally. It is essentially negligible (i.e., smaller than
the statistical errors) outside the Galactic plane ($|b| > 10\degr$)
where the diffuse emission is faint and varying on large angular scales.
It is also not much of a worry in the high band ($>$ 1~GeV) where
the PSF is sharp enough that the bright sources dominate the background
under the PSF.
But it is a serious issue inside the Galactic plane
{\bf ($|b| < 10\degr$)}
in the low band ($<$ 1~GeV) and particularly
inside the Galactic ridge ($|l| < 60\degr$) where the diffuse emission
is strongest and very structured, following the molecular cloud distribution.
It is not easy to assess precisely how large the uncertainty is, for lack
of a proper reference. We have tried re-extracting the source fluxes
assuming a very different diffuse model, and the results tend to show that
the systematic uncertainty more or less follows the statistical one
(i.e., it is larger for fainter sources) and is of the same order.
We have not increased the errors accordingly, though, because this
alternative model does not fit the data as well as the reference model
where the differences in the source fluxes
are largest.

The net result of these considerations is that
we expect our high energy fluxes to be reasonably accurate,
but the low energy fluxes are not as reliable and should be treated with
particular caution in the Galactic ridge.

\section{Source Association and Identification}
\label{run:assoc}
Even with the superior angular resolution of LAT compared to previous-generation $\gamma$-ray telescopes, the source location accuracy is not  good enough to draw firm conclusions based on positional coincidence in most cases.  A typical LAT error circle contains multiple stars, galaxies, X-ray sources, infrared sources, and radio sources. Determination of the nature of a given LAT source must therefore rely on {\bf more information than only simple location}.  Two principles lead the search for counterparts:

\begin{itemize}
\item Variability is a powerful diagnostic, particularly considering that many $\gamma$-ray sources are known to be variable.  Searches for periodic variability (such as rotational and orbital motion) offer opportunities for unique identifications.  Determining variability correlated with that seen at other wavelengths is another approach.
\item LAT $\gamma$-ray sources are necessarily nonthermal objects involving large energy transfers.  Physical properties of any candidate counterpart must be consistent with generation of a significant luminosity of gamma radiation. 
\end{itemize}

In this analysis, the LAT team makes a clear distinction between a source {\it identification} and an {\it association} with an object at another wavelength.  A firm identification of a source is based on a timing characteristic such as a periodicity for a pulsar or binary or a variability correlated with observations at another wavelength in the case of a blazar. An association is made for a statistically-improbable positional coincidence of a plausible $\gamma$-ray-producing object with a LAT source.  

\subsection{Automated Source Associations}

In anticipating the large number of $\gamma$-ray sources that will be detected by the LAT
in the course of the mission, we implemented an automated source association pipeline
that attempts to make quantified associations between LAT sources and potential counterparts.
In its implementation for the Bright Gamma-Ray Source List the pipeline is almost
exclusively based on positional coincidence, yet is driven by past knowledge about
GeV source classes (pulsars and blazars) and physical expectations {\bf (such as total luminosity and nonthermal emission implying particle acceleration).} 
Future implementations will also include figure-of-merit (FoM) approaches \citep{SowardsEmmerd2003}
but these first require careful training on firmly identified source classes.

For each LAT source the probability of association with a source in the counterpart catalog
is estimated using a Bayesian approach \citep[e.g.,][]{deRuiter1977,Sutherland1992} that considers the spatial match between
LAT source and counterpart in light of the position uncertainty $r_{95}$ and the chance
coincidence probability as inferred from the local source density in the counterpart catalog.
Specifically we calculate the posterior probability of association
\begin{equation}
{\rm P}_{\rm post} = \left( 1 + \frac{1 - {\rm P}_{\rm prior}}{{\rm P}_{\rm prior}} 
                                          \frac{\pi \rho \, r_{95}^2}{2.996} \, e^\Delta \right)^{-1}
\end{equation}
where
{\bf
$\rho$ is the local counterpart density,
$\Delta = 2.996 \times r^2 / r_{95}^2$,
$r$ is the angular separation between LAT source and catalog counterpart, and
${\rm P}_{\rm prior}$ is the prior probability of association that we use here as a
constant tuning parameter whose value is adjusted for each counterpart catalog to give an approximately constant false association rate among the catalogs considered.
Since the value of ${\rm P}_{\rm post}$ depends on the choice of ${\rm P}_{\rm prior}$
we arbitrary define a counterpart as a possible association if 
${\rm P}_{\rm post} \ge 0.5$.
We applied our pipeline to random realizations of plausible LAT catalogs
\footnote{
The plausible LAT catalogs contained 1000 sources of which 75\% were
distributed isotropically over the sky and 25\% were distributed along the
Galactic plane following a 2D Gaussian shaped density profile with
$\sigma=40\deg$ in longitude and $\sigma=2\deg$ in latitude.
For each source an error radius $r_{95}$ of $0.2\deg$ has been assumed.
}
in order to find for each counterpart catalog the value of ${\rm P}_{\rm prior}$ that 
does not produce more than a single spurious association.
}

Table \ref{tab:catalogs} summarizes the catalogs that have been used in our automatic
association procedure.
We also quote the prior probabilities that have been employed and
give the total number of objects in each catalog.
Note that we make an exception to our procedure when we cross-correlate the
EGRET 3EG and EGR and AGILE AGL catalogs with the LAT sources.
Since in these cases the uncertainties in the localization of the counterparts is worse than
for the LAT sources we consider all EGRET and AGILE sources as possible counterparts
if the LAT and counterpart separation is less than the quadratic sum of their $95\%$ confidence error radii.
\begin{table}[!t]
  \footnotesize
  \caption{\label{tab:catalogs}
  Catalogs used for automatic source association.
  For clarity the table has been divided into Galactic, extragalactic and $\gamma$-ray source
  catalogs.
  }
  \begin{flushleft}
	\begin{tabular}{lrlcl}
	\hline
	\hline
	\noalign{\smallskip}
	Name & Objects & ${\rm P}_{\rm prior}$ & Selection & Reference \\
	\noalign{\smallskip}
	\hline
	\noalign{\smallskip}
	Pulsars (high $\dot{E}/d^2$) & 100 & 0.29 & $\dot{E}/d^2 > 5\,10^{33}$ erg cm$^{-2}$ s$^{-1}$ & \cite{ATNFcatalog} \\
	Pulsars (low $\dot{E}/d^2$) & 1527 & 0.044 & $\dot{E}/d^2 \le 5\,10^{33}$ erg cm$^{-2}$ s$^{-1}$ & \cite{ATNFcatalog} \\
	PWN & 69 & 0.5 & & \cite{PWNcatalog}\tablenotemark{a} \\
	SNR & 265 & 0.033 & & \cite{SNRcatalog}\tablenotemark{b} \\
	HMXB & 114 & 0.17 & & \cite{HMXBcatalog} \\
	LMXB & 187 & 0.19 & & \cite{LMXBcatalog} \\
	Microquasars & 15 & 0.5 & & \cite{MQOcatalog} \\
	Globular clusters & 147 & 0.5 & & \cite{GlobClusterCatalog} \\
	\noalign{\smallskip}
	\hline
	\noalign{\smallskip}
	Blazars (CGRABS) & 1625 & 0.14 & & \cite{CGRaBS} \\
	Blazars (BZCAT) & 2686 & 0.043 & &  \cite{BZcatalog} \\
	Flat Spectrum Radio Sources (CRATES) & 10272 & 0.022 & &  \cite{CRATES} \\
	\noalign{\smallskip}
	\hline
	\noalign{\smallskip}
	3EG catalog & 271 & n.a. & &  \cite{3rdCat} \\
	EGR & 189 & n.a. & &  \cite{EGR} \\
	AGL & 40 & n.a. & &  \cite{AGILE} \\
	\noalign{\smallskip}
	\hline
	\end{tabular}
	\tablenotetext{a}{http://www.physics.mcgill.ca/~pulsar/pwncat.html}
	\tablenotetext{b}{http://www.mrao.cam.ac.uk/surveys/snrs/}
  \end{flushleft}
\end{table}

The pulsar catalog (the ATNF Pulsar Catalogue, \citealt{ATNFcatalog}) is special in that we split it into high and low $\dot{E}/d^2$ 
sub-samples, where $\dot{E}$ is the rate of energy loss of the pulsar and d is the distance.
High $\dot{E}/d^2$ has been proposed as a good estimator of a pulsar's $\gamma$-ray
visibility \citep{DAS08}, and downselecting the catalog to the 100 best candidates
allows for a relatively large prior probability without inflating the number of false
positives.
For the remaining pulsars {\bf our Monte Carlo simulations required a much smaller
prior probability} (to keep
the chance coincidences low) at the expense of reducing the number of potential
associations.
This procedure can be considered as a simple binary figure-of-merit approach
which favours revealing high $\dot{E}/d^2$ counterparts of LAT sources.

The performance of our association scheme is illustrated in Figure 9, which
shows the distribution of normalized angular separations between LAT sources and 
counterparts; the normalization is done with respect to the measured localization
uncertainty (Sect.~\ref{run:localization}). 
We also show the expected distribution for the case that all sources have been correctly 
associated (dotted line) or are spurious associations (dashed line).
Obviously, the observed distribution clearly follows the first trend, suggesting that most of our associations
are indeed reasonable and that our efforts to reduce the number of false positives were
successful.
We notice that the histogram shows a slight trend to smaller angular separations than
expected, which might result from a slight overestimation of our source localization
uncertainties.

\begin{figure}
\label{fig9}
\centering
\plotone{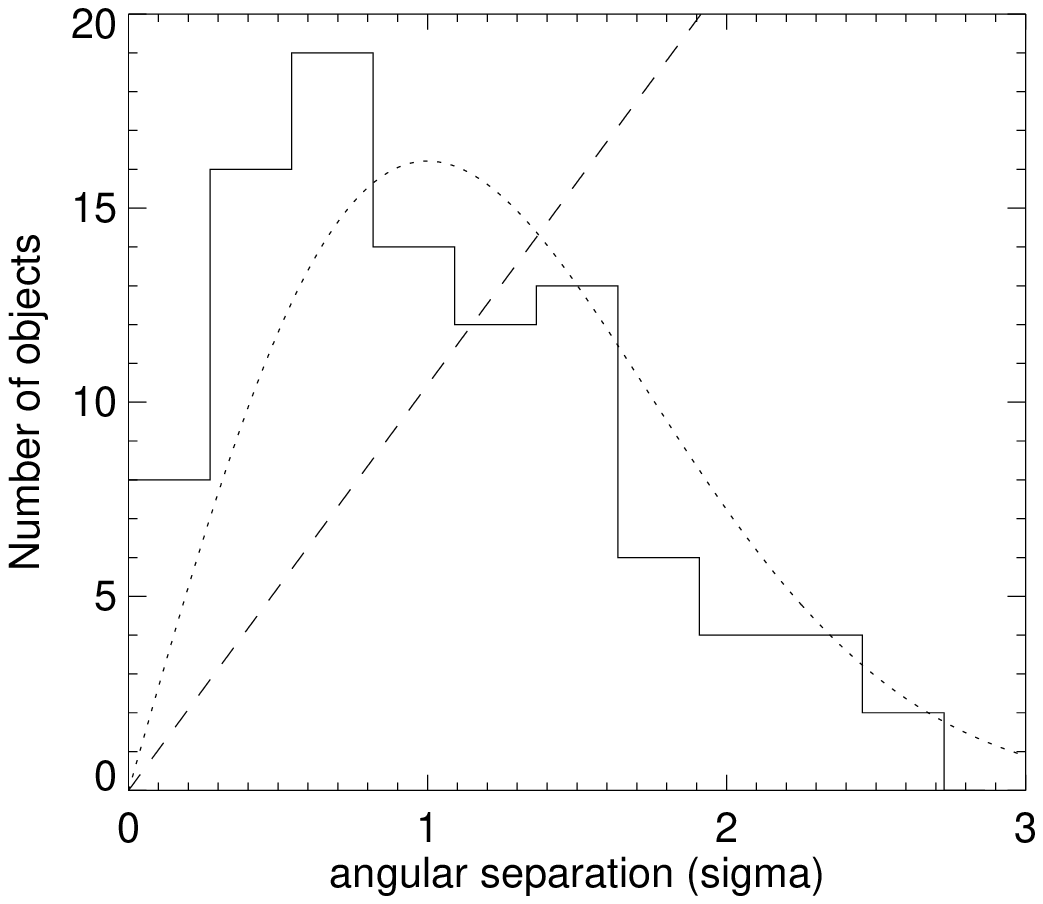}
\caption{Distribution of angular separations between LAT sources and counterpart catalog
associations expressed as $\sigma=0.405 \, r/r_{95}$.
The expected distribution in the case that all sources have been correctly associated
is given as dotted line. {\bf The peak at somewhat lower angular separation than the dotted prediction might indicate slightly better position determinations on average than (conservatively) assumed in this paper.}
Conversely, the expected distribution in the case that all sources are spurious associations
is given as the dashed line. 
}
\end{figure}

\subsection{Alternate Associations, Firm Identifications, and Special Cases}

\subsubsection{Active Galactic Nuclei (AGN)}

Active Galactic Nuclei have been recognized since the EGRET era as a well-defined class of gamma-ray sources.  For this reason, we have adopted an alternate method of finding AGN associations beyond the automated association procedure described in the previous section.  This method uses a Figure of Merit (FoM) approach similar to the one described by \cite{CGRaBS}, based not only on positional proximity but also on radio spectral index, X-ray flux, and radio flux  \citep{AGNpaper}.  Details of this association procedure, including the calculated probabilities from {\bf both the FoM and automated association approaches, can be found in that paper.  Although most of the associations are found by both methods, about 11\% are found only by one of the two. In order to maintain consistency with the LAT AGN paper  \citep{AGNpaper}, we show any association found by either method. }

It should be emphasized, however, that \cite{AGNpaper} chose to  apply the AGN analysis only to parts of the sky with Galactic latitudes more than 10$^\circ$ from the plane in order to have a more uniform sample, while the present analysis covers the entire sky.  AGN are seen by LAT at lower Galactic latitudes, because the Galaxy is largely transparent to $\gamma$- rays.  Due to Galactic extinction and source confusion, AGN identification is more difficult at low latitudes. Some of the unassociated LAT sources in this part of the sky can be expected to have AGN counterparts in further analysis, which is beyond the scope of this paper. 

\subsubsection{Firm identifications}
For this early source list from the LAT, we have taken the conservative view that association, even with high probability, is not equivalent to firm identification. Error circles are still large compared to source localization at longer wavelengths.  We adopt the approach that firm identification for the 0FGL sources is limited to those for which variability can unambiguously establish the source.  

Firm identifications of pulsars are based on seeing the pulsations in the $\gamma$-ray data
with high confidence.
Using several statistical tests, we require that the $\gamma$-ray distribution in pulsar phase be inconsistent
with random at a probability level of $10^{-6}$ or smaller.
Examples are the 6 pulsars confirmed from the EGRET era, the radio-quiet pulsar found in the CTA~1
supernova remnant \citep{CTA1}, 
PSR~J0030+0451  \citep{PSR0030},  
PSR~J1028$-$5819  \citep{PSR1028}
and PSR J2021+3651 \citep{PSR2021LAT}.
In total the 0FGL source list includes  30 firm pulsar identifications. One third of the sources within 10$^\circ$ of the Galactic Plane have now been identified with pulsars. 

The HMXB system LSI +61 303 is firmly identified based on the observation of the orbital period of the binary system \citep{LSI+61}.  A search for periodicity in the similar source LS5039 is still in progress. 

Firm identifications of AGN depend on finding correlated multiwavelength activity.
This work is ongoing.

\subsubsection{Special cases -- Pulsar Wind Nebulae (PWNe) and Supernova Remnants (SNRs)}

SNRs and PWNe that are positionally correlated with LAT sources are not listed
as individual associations in the main source list table.  Statistical
indications are that SNRs that were coincident with EGRET sources are
significantly correlated with the 0FGL sources.  However, the large number of
pulsars detected by the LAT, including radio-quiet pulsars \citep[e.g.,][]{CTA1}, suggests that even a positional coincidence with an SNR of an age,
distance, and environment plausible for a $\gamma$-ray source may be due
to a $\gamma$-ray pulsar.  Of the 0FGL sources positionally associated with
PWNe or SNRs, approximately 40\% have already been found to contain
$\gamma$-ray pulsars. At the present level of sensitivity for the LAT-detected
pulsars, only the Crab has shown evidence for off-pulse emission that can be
attributed to a PWN or SNR. Until the possibility of pulsed emission for such
sources can be ruled out, we are reluctant to make any claims about individual
PWNe or SNRs as possible LAT detections. Effectively, the high rate of pulsar
detections increases the burden of proof for PWN and SNR candidates, for
example via studies of source extents.

Table 2 shows the 0FGL sources that are associated positionally with PWNe and
SNRs, plus 4 pulsars that do not (yet) show evidence of $\gamma$-ray
pulsation.  \citet{Torres2003} considered several of the SNRs in Table 2 in
terms of their potential to be $\gamma$-ray counterparts to unidentified
low-latitude EGRET sources.  In the case of SNR G284.3$-$1.8 they argued that
PSR J1013$-$5915 was more probably the $\gamma$-ray source.  

\begin{deluxetable}{lrrl}
\setlength{\tabcolsep}{0.04in}
\tablewidth{330pt}
\tabletypesize{\scriptsize}
\tablecaption{Potential Associations for Sources Near SNRs and
PWNe\label{table_aux_assoc}}
\tablehead{
\colhead{Name 0FGL} &
\colhead{$l$} &
\colhead{$b$} &
\colhead{Assoc.}
}
\startdata
 J0617.4+2234 & 189.08 & 3.07 & SNR G189.1+3.0 (IC 443)   \\
 J1018.2$-$5858 & 284.30 & $-$1.76 & SNR G284.3-1.8 (MSH 10-53), PSR
J1013$-$5915   \\
 J1106.4$-$6055 & 290.52 & $-$0.60 & SNR G290.1$-$0.8 (MSH 11-61A), PSR
J1105$-$6107   \\
 J1615.6$-$5049 & 332.35 & $-$0.01 & SNR G332.4+0.1, PWN G332.5-0.28, PSR
B1610-50   \\
 J1648.1$-$4606 & 339.47 & $-$0.71 & PSR J1648$-$4611   \\
 J1714.7$-$3827 & 348.52 & 0.10 & SNR G348.5+0.1   \\
 J1801.6$-$2327 & 6.54 & $-$0.31 & SNR G6.4$-$0.1  (W28) \\
 J1814.3$-$1739 &  13.05 & $-$0.09 & PWN G12.82$-$0.02   \\
 J1834.4$-$0841 & 23.27 & $-$0.22 & SNR G23.3$-$0.3 (W41)  \\
 J1855.9+0126 &  34.72 & $-$0.35 & SNR G34.7$-$0.4 (W44)  \\
 J1911.0+0905 &  43.25 & $-$0.18 & SNR G43.3$-$0.2   \\
 J1923.0+1411 & 49.13 & $-$0.40 & SNR G49.2$-$0.7 (W51)   \\
 J1954.4+2838 &  65.30 & 0.38 & SNR G65.1+0.6   \\
\enddata
\tablecomments{See text, Sect. 4.2.3.  These sources are marked with a
$\dagger$ in Table 6. They may be pulsars rather than the SNR or PWN named.}
\end{deluxetable}

\section{The Source List}
\label{run:sourcelist}

The bright source list is presented as a single table (Table 6). Table 3 is a description of the columns in Table 6. Within the table, sources that have firm identifications or tentative associations are listed by class.  Table 4 describes those classes. Figure 10 shows the locations of the 205 bright sources, in Galactic coordinates.  All associations with specific source classes are also shown.  Figure 11 is an enlargement of the bright source map, showing the region of the inner Galaxy. This list is available as a FITS file from the Fermi Science Support Center.

\begin{deluxetable}{ll}
\setlength{\tabcolsep}{0.04in}
\tabletypesize{\scriptsize}
\tablecaption{LAT Bright Source List Description\label{table3}}
\tablehead{
\colhead{Column} &
\colhead{Description} 
}
\startdata
Name &  0FGL JHHMM.m+DDMM, constructed according to IAU Specifications for Nomenclature; m is decimal\\
  &  minutes of R.A.; in the name R.A. and Decl. are truncated at 0.1 decimal minutes and 1\arcmin, respectively \\
R.A. & Right Ascension, J2000, deg, 3 decimal places  \\
Decl. & Declination, J2000, deg, 3 decimal places \\
 $l$  & Galactic Longitude, deg, 3 decimal places \\
 $b$  & Galactic Latitude, deg, 3 decimal places \\
 $\theta_{95}$ & Radius of 95\% confidence region, deg, 3 decimal places\\
 $TS^{1/2}$ & Square root of likelihood $TS$ from 200~MeV--100~GeV analysis, used for the $TS >$ 100 cut, 1 decimal place \\
 $F_{23}$ & Flux 100~MeV$-$1~GeV (i.e., log$_{10}$E = 2$-$3), 10$^{-8}$ cm$^{-2}$ s$^{-1}$, 2 decimal places \\
 $\Delta F_{23}$  & 1-$\sigma$ uncertainty on $F_{23}$, same units and precision. A 0 in this column indicates that the entry in the $F_{23}$ flux column is an upper limit.  \\
 $TS_{23}^{1/2}$ & Square root of $TS$ for the 100~MeV--1~GeV range, 1 decimal place\\
 $F_{35}$ &  Flux for 1~GeV--100~GeV (i.e., log$_{10}$E = 3$-$5),10$^{-8}$ cm$^{-2}$ s$^{-1}$, 2 decimal places \\
 $\Delta F_{35}$ & 1-$\sigma$ uncertainty on $F_{35}$, same units and precision \\
 $TS_{35}^{1/2}$  & Square root of $TS$ for the 1~GeV$-$100~GeV range, 1 decimal place \\
 Var. & T indicates $<$~1\% chance of being a steady source on a weekly time scale; see Sect.~\ref{run:variability}  \\
 $\gamma$-ray Assoc.  & Identification or positional associations with 3EG, EGR, {\bf or AGILE} sources \\
 Class & Like 'ID' in 3EG catalog, but with more detail (see Table 4).  {\bf Capital letters indicate firm identifications; lower-case letters indicate associations.} \\
 ID or Assoc.  & Identification or positional associations with potential counterparts \\
 Ref. & Reference to associated paper(s),  \\  
\enddata
\end{deluxetable}

\begin{deluxetable}{ll}
\setlength{\tabcolsep}{0.04in}
\tabletypesize{\scriptsize}
\tablecaption{LAT Bright Source List Source Classes\label{table4}}
\tablehead{
\colhead{Class} &
\colhead{Description} 
}
\startdata
 PSR & pulsar \\
 pwn & pulsar wind nebula \\
 hxb & high-mass X-ray binary (black hole or neutron star) \\
 bzb & BL Lac type of blazar \\
 bzq & FSRQ type of blazar \\
 bzu & uncertain type of blazar \\
 rdg & radio galaxy \\
 glb & globular cluster \\
 $\dagger$ & Special case - potential association with SNR or PWN (see Table 2) \\
\enddata
\tablecomments{Designations shown in capital letters are firm identifications; lower case letters indicate associations. In the case of AGN, many of the associations have high confidence \citep{AGNpaper}. Among the pulsars, those with names beginning with LAT are newly discovered by the LAT. }
\end{deluxetable}

\begin{figure}
\epsscale{.80}
\plotone{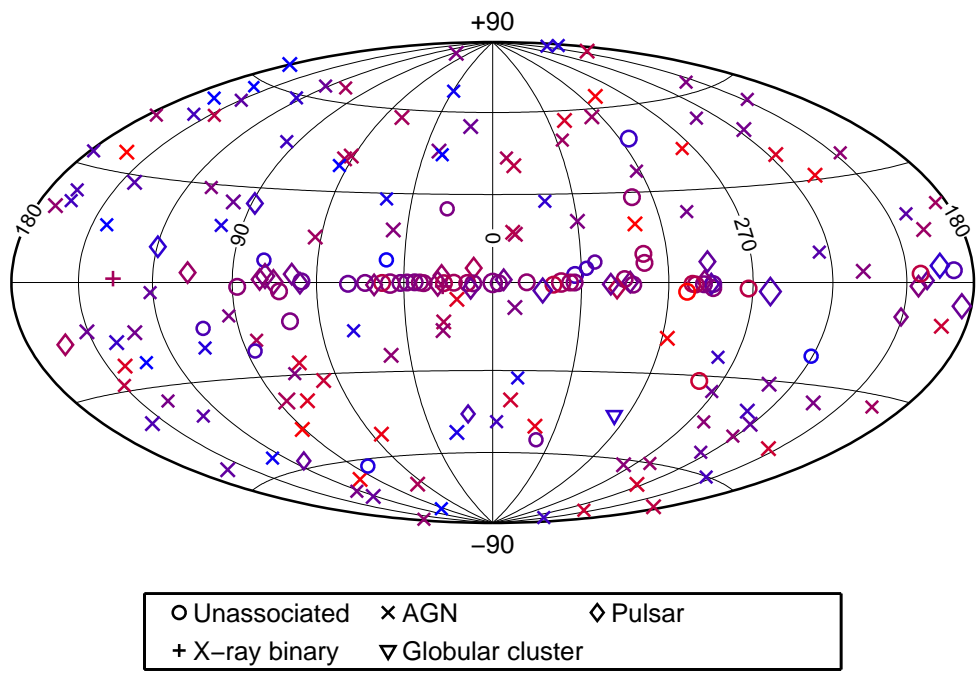}
\caption{{The LAT Bright Source List, showing the locations on the sky (Galactic coordinates in Aitoff projection) coded according to the legend. {\bf Although quantitative spectral information is not presented, the colors of the symbols indicate relative spectral hardness on a sliding scale. Symbols more blue in color indicate sources with harder spectra than those that are more red.}}\label{fig10}}
\end{figure}

\begin{figure}
\epsscale{.10}
\includegraphics[ scale=0.8]{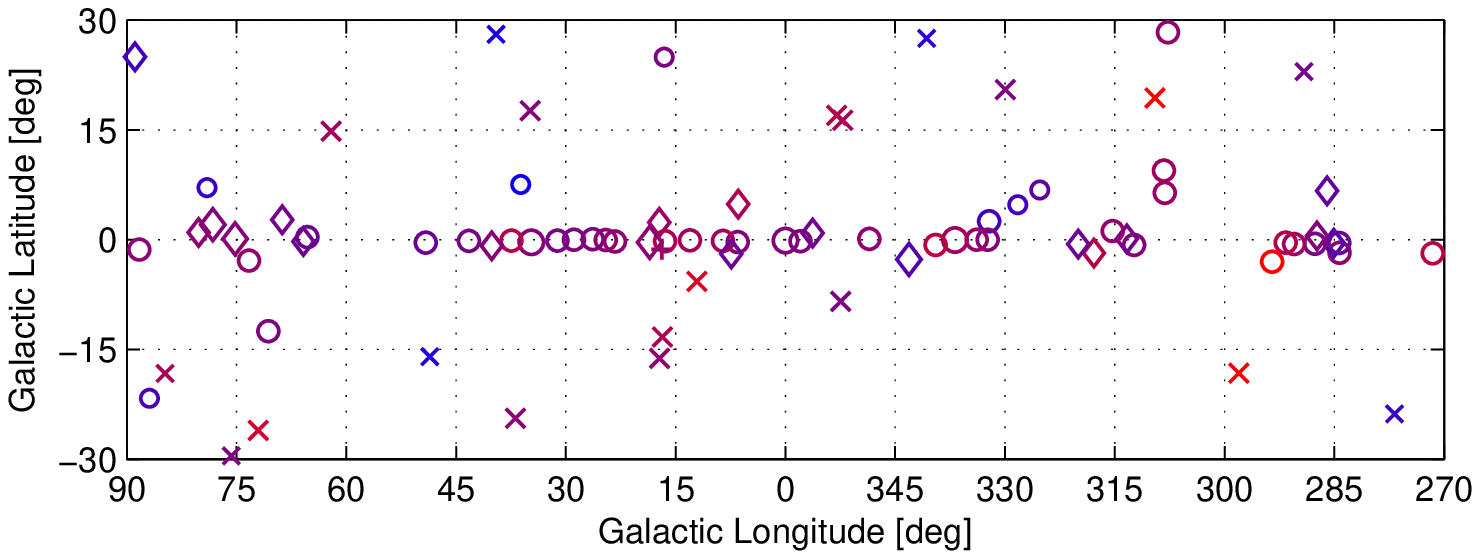}
\caption{{The LAT Bright Source List, showing the locations of sources in the inner Galaxy. The legend is the same as in Figure 10.}\label{fig11}}
\end{figure}

\section{Discussion}

As is clear from the references in this paper, much of the work on the early data from $Fermi$ LAT is still in progress.  In particular, we re-emphasize several caveats for use of this bright source list:

\begin{itemize}

\item Ongoing efforts to understand the calibration and improve the analysis techniques are underway.  In many respects, therefore, the 0FGL source list is quite preliminary.  Significant improvements are expected before the construction of the first full LAT catalog.  

\item The GALPROP diffuse model used in the analysis is still evolving.  Matching the model to the large-scale emission is an iterative process.  The diffuse model is particularly important for sources near the Galactic Plane. 

\item This source list is limited to high-confidence detections.  It is not a full catalog.

\item The 0FGL list information in two broad energy bands is not appropriate for detailed spectral modeling. 

\item This work is a ``snapshot'' of the LAT results covering only the {\bf observation} time period August to October, 2008.

\item The use of the diffuse class of events means that LAT has little sensitivity below 200~MeV for this particular analysis.  As noted by \cite{AGNpaper} in their analysis of AGN, the LAT is more sensitive to hard-spectrum sources than previous satellite instruments. 

\end{itemize}

Despite these issues, the present work demonstrates the power of the LAT to make high-energy $\gamma$-ray observations and shows its potential for future discoveries.  Although we feel it premature to draw far-reaching conclusions, some results stand out. 

  \begin{figure}
\includegraphics[ scale=1.0]{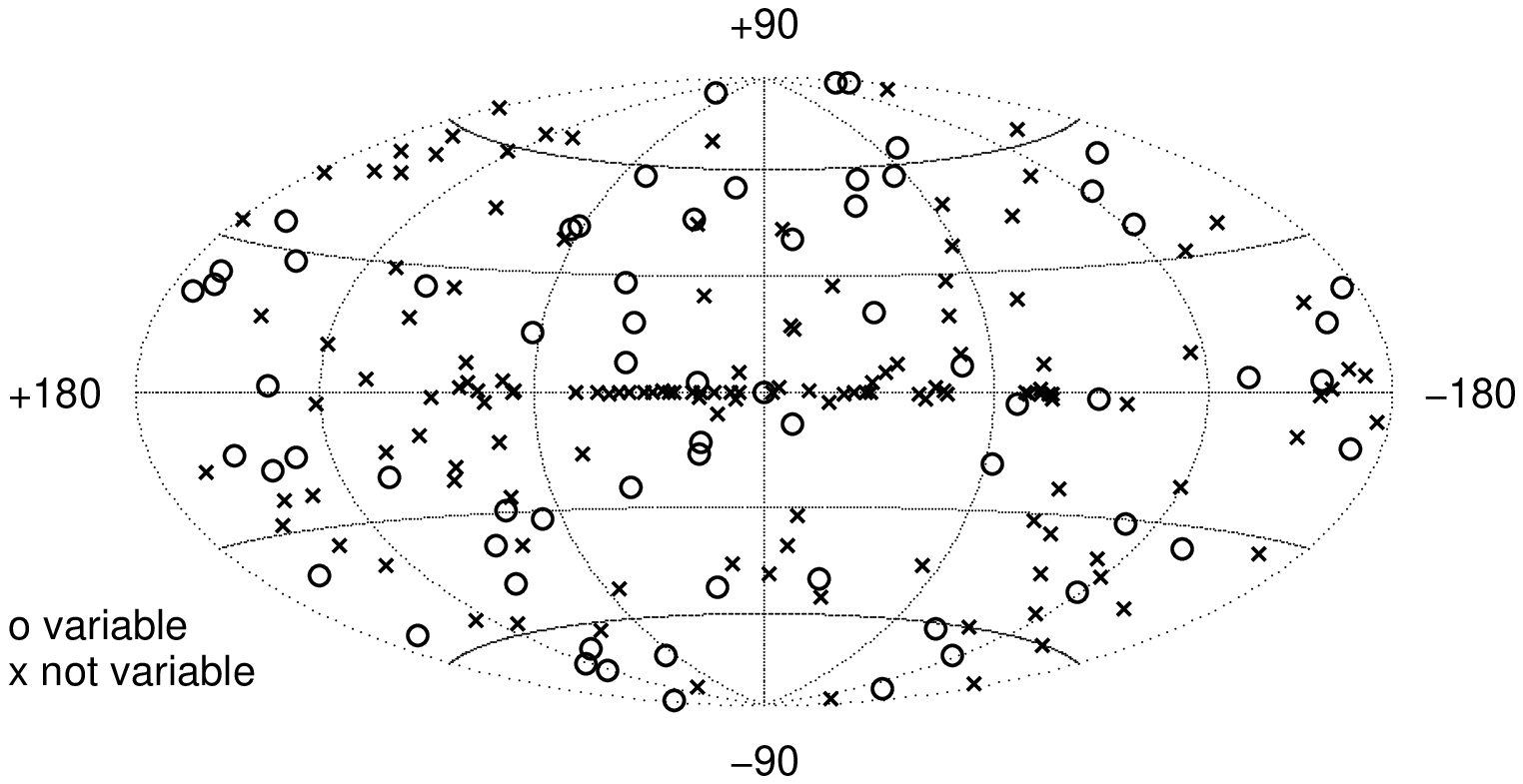}
\caption{{Locations of variable ($circles$) and non-variable ($crosses$) 0FGL sources, using the definition of variability in Sec. 3.5. The analysis is sensitive to variations on time scales of weeks to $\sim$2 months.}.\label{fig12}}
\end{figure}

\subsection{Characteristics of the 0FGL Sources}

\begin{itemize}
    
  \item Both Galactic and extragalactic populations are visible.  73 sources are found within 10$^\circ$ of the Galactic Plane, where they exhibit a characteristic concentration in the inner Galaxy; 132 are seen at higher Galactic latitudes. 
  
   \item 66 of the bright LAT sources show solid evidence of variability {\bf on weekly time scales during this three-month interval}.  Figure 12 shows the locations of variable and non-variable sources, in Galactic Coordinates. 
   
   \item The typical error radii for the sources (95\% confidence) are less than 10\arcmin. 
   
   \item The Galactic latitude distribution of unassociated/unidentified $\gamma$-ray sources is very narrow (FWHM $<$0.5$^\circ$). If we assume a scale height for a Galactic population of 40 pc \citep{Guibert78}, such a narrow latitude distribution points to a Galactic $\gamma$-ray source population with average distance in excess of  40/sin0.5$^\circ$, namely 4.5 kpc. 
   
   \end{itemize}

\subsection{Comparisons with Other High-Energy $\gamma$-ray Results}

Before $Fermi$, the EGRET results represented the most complete view of the high-energy sky, but those results applied to the 1991-2000 era.  In light of the variability seen in the EGRET $\gamma$-ray sources, significant differences were expected.  A contemporaneous mission to $Fermi$ is AGILE, which began operations over a year before $Fermi$ and continues to operate.  Here is a summary of comparisons with these missions:


\begin{itemize}
    
  \item Of the 205 0FGL sources, 60 have nearby counterparts (the LAT source 95\% uncertainty overlapping that of the  EGRET source) found by the automated analysis in the 3EG  catalog (271 sources); 43 in the EGR catalog (189 sources).   Most of the sources seen by EGRET in the 1990s were not seen by LAT as bright sources in 2008.  Approximately 40\% of the bright LAT sources off the plane that have no former EGRET counterparts are found to be variable.  
  
  \item {\bf EGRET found few sources with flux less than 10 $\times$ 10$^{-8}$ photons(E$>$100~MeV) cm$^{-2}$ s$^{-1}$.  A number of the 0FGL sources have fluxes well below this value (e.g., 0FGL J0033.6$-$1921).  Such sources would not have been visible to EGRET.} 
  
  \item {\bf Some sources, such as 0FGL J0428.7$-$3755, associated with blazar PKS 0426$-$380, have flux values well above the EGRET threshold but were not seen by EGRET and yet are not noted as being variable in the 0FGL data.  Such sources serve as a reminder that blazars are variable on many time scales, and the 0FGL sample covers only three months.}  
  
  \item Considering the highest-confidence sources, in its lifetime EGRET found 31 sources (in either the 3EG or EGR catalogs or both) with confidence level of $>$10-$\sigma$.  The 0FGL list shows the dramatic improvement in sensitivity of the LAT.

\item Five of the EGRET sources seen at 10-$\sigma$ significance (all associated with flaring blazars: NRAO 190, NRAO 530, 1611+343, 1406$-$076, and 1622$-$297) do not appear in the LAT bright source list.   

\item 28 of the EGRET sources that have counterparts in the 0FGL list were previously listed as unidentified.  Half of these, 14, have now been firmly identified in this early LAT analysis. 13 are pulsars; 1 is a HMXB. 
  
  \item Of the 40 sources in the first AGILE catalog (which is contemporaneous but does not overlap in time with the 0FGL data), 32 are also found in the 0FGL list and 7 more, while not formally overlapping, are ``near misses'' to 0FGL sources.  The one exception is AGL J1238+04, a source associated with a FSRQ.  A LAT source consistent in position with this one is found at a lower significance during the first 3 months of operation, but the source has since flared \citep{ATEL1239}.  
      
\end{itemize}

\subsection{Some Results from the Association Analysis} 

Table 5 summarizes the census of associations in the Bright Source List.  {\bf The numbers of these associations that are considered firm identifications are shown in parentheses.} 

\begin{deluxetable}{lr}
\setlength{\tabcolsep}{0.04in}
\tabletypesize{\scriptsize}
\tablecaption{LAT  Bright Source List Source Associations (Firm Identifications)\label{table5}}
\tablehead{
\colhead{Class} &
\colhead{Number} 
}
\startdata
 Radio/X-ray pulsar (PSR) & {\bf 15 (15)}\\
 LAT gamma-ray pulsar (LAT PSR) & {\bf 15 (15)} \\
HMXB & 2 (1) \\
 BL LAC (bzb) & 46 (0)\\
 FSRQ (bzq) & {\bf 64 }(0) \\
Other blazar (Uncertain type, bzu) &{\bf 9} (0) \\
 Radio galaxy (rdg) & 2 (0)\\
 Globular Cluster (glb, see text) & 1 (0) \\
 LMC (see text) & 1 (0) \\
 $\dagger$ Special cases (see Table 2) & {\bf 13} (0) \\
 Unassociated & {\bf 37} (0) \\
\enddata
\end{deluxetable}

\begin{itemize}
  \item The AGN class (121 members) is the largest source type associated in the LAT data.  Details of the analysis, together with the implications for AGN studies, are given by \cite{AGNpaper}.  Two of the AGN found in this analysis are associated with radio galaxies; the rest are categorized as blazars.  Note that 5 of the 0FGL AGN are not included in the \cite{AGNpaper} analysis, because they are found within 10$^\circ$ of the Galactic Plane. 
  
  \item Pulsars, including young radio pulsars, millisecond radio pulsars, and radio-quiet pulsars, form another well-defined class  (30 members) in the LAT bright source list.  
  
  \item Among the 0FGL sources, no associations were found with LMXB, starburst galaxies, prominent clusters of galaxies or Seyfert galaxies.
  
  \item 2 associations were found with HMXB sources, both of which are also seen at TeV energies:  LSI +61 303 \citep{LSIMAGIC} and LS 5039 \citep{LS5039HESS}.  The association of 0FGL J0240.3+6113 with LSI +61 303 is considered a firm identification based on the orbital periodicity seen in the LAT emission \citep{LSI+61}.  Analysis of LS 5039 is in progress. 
    
  \item Globular cluster NGC 104 = 47 Tuc is associated with LAT source 0FGL J0025.1$-$7202; it should be emphasized that this globular cluster contains at least 23 millisecond radio pulsars and presumably contains many more as-yet undetected neutron stars. 
  
  \item 0FGL J0538.4$-$6856 is seen in the direction of the Large Magellanic Cloud.  The LMC X-ray pulsar associated with this source by the automated software (PSR J0537$-$6910) is one possibility.  The source is also consistent with the direction of the 30 Doradus star-forming region.  Work on this part of the sky is still in progress. 
    
  \item 0FGL J0617.4+2234 lies within the projected direction of the shell of SNR IC443. A TeV source has also been seen close to the position of the LAT source.  Detailed analysis of the LAT source is in progress. 
  
\item 0FGL J0910.2$-$5044, although visible in the summed map, was seen primarily as a Galactic transient in October, 2008 \citep{ATel.1788}. 
  
  \item 0FGL J1746.0$-$2900 lies close to the Galactic Center.  Modeling the diffuse emission in this general region is challenging.  We consider any conclusions about the association of this source with the Galactic Center or other candidate $\gamma$-ray emitters in this region to be premature.  The variability flag for this source is True, but the source is not extremely variable.  This source barely met the criterion for being called variable. Work on this region is in progress. 
  
   \item {\bf 37} of the 0FGL sources have no obvious counterparts at other wavelengths.

  \end{itemize}

\subsection{TeV Comparisons}

Associations with TeV sources are based in this work only on positional correlation.  Physical modeling or correlated variability would be needed in order to draw any conclusions from these associations.  This is not an exhaustive list.  We have omitted associations with blazars,  well-known objects such as the Crab Nebula, and sources discussed previously in the text. 
 
\begin{itemize} 

 \item 0FGL J1024.0$-$5754
is spatially consistent with HESS J1023$-$575, itself not yet firmly identified, but noted for its possible connection to the young stellar cluster Westerlund 2 in the star-forming region RCW49 \citep{Westerlund2TEV}.

 \item 0FGL J1418.8$-$6058
is spatially coincident with  HESS J1418$-$609 \citep{2006HESS1418}, which may be the PWN powered by the LAT-discovered LAT PSR J1418$-$60. 
 
 \item 0FGL J1615.6$-$5049
is spatially coincident with HESS J1616-508, which has been suspected being the PWN of PSR J1617$-$5055 \citep{PSR1617PWN}, \citep{PSR1617TEV}.   

\item OFGL J1741.4-3046 is spatially consistent with the unidentified HESS J1741-302. \citep{Tibolla2008}.  See the note in the previous section about LAT analysis in the Galactic Center region.  

 \item 0FGL J1805.3$-$2138
is spatially coincident with HESS J1804$-$216 \citep{2005HESSPlane}. Formally still unidentified, HESS J1804$-$216 has been noted for possible counterparts in SNR G8.7$-$0.1, W30, or PSR J1803$-$2137. At this stage, we are not able to make a firm identification of the LAT source  with any of the counterpart hypotheses.

 \item 0FGL J1814.3$-$1739
is spatially coincident with HESS J1813$-$178  \citep{2005HESSPlane}, which has been classified as a composite SNR, characterized by a shell-type SNR with central PWN candidate, not to be distinguishable given the angular resolution of present VHE observatories. At this stage, we are not able to settle either on a SNR or on a PSR/PWN scenario for connecting HESS J1813$-$178 with the LAT source, leaving this study to a follow-up investigation.

 \item 0FGL J1834.4$-$0841
is spatially coincident with HESS J1834$-$087  \citep{2005HESSPlane}. Formally still unidentified, HESS J1834$-$087 was proposed to be explained in emission scenarios involving SNR W41, hadronic interactions with a giant molecular cloud, and/or PSR J1833$-$0827.  See Table 2. PSR J1833$-$0827 is not consistent in position with the LAT source. At this stage, we are not able to draw conclusions on a possible connection of the LAT source to the presented counterpart hypothesis.

\item OFGL J1923.0+1411 is spatially coincident with HESS J1923+141, which is also spatially consistent with SNR G49.2-0.7 (W51).  See Sec. 4.2.3 and Table 2 for a discussion of LAT source associations with SNRs and PWNe.

  \item 0FGL J2032.2+4122, conclusively identified as a PSR, 
is spatially coincident with TeV 2032+4130 seen by HEGRA \citep{HEGRA} and Milagro source MGRO J2031+41\citep{Milagro}. Formally still unidentified, TeV J2032+4130 was noted for being close to the direction of the massive stellar cluster association Cygnus OB2.  MGRO J2031+41, also unidentified, was reported as an extended and possibly confused source that could only be explained in part by the emission from TeV 2032+4130.  We leave the possible association of LAT PSR J2032+41 with TeV J2032+4130 or MGRO 2031+41 to a detailed subsequent study.

 \item Finally, it is noteworthy that LAT pulsars are found also in the error circles of four Milagro detected or candidate sources \citep{Milagro}:
 \begin{itemize}
 \item 0FGL J2020.8+3649 (MGRO 2019+37)
 \item 0FGL J1907.5+6002 (MGRO 1908+06)
 \item 0FGL J0634.0+1745 (C3 - candidate, Geminga)
 \item 0FGL J2229.0+6114 (C4 - candidate)
 \end{itemize} 

\end{itemize}

Detailed discussion of individual sources is beyond the scope of this paper.  By noting these positional coincidences, we call attention to areas of work still in progress on the $Fermi$ LAT data.  The 0FGL list is a starting point for additional research in many areas. 

\acknowledgments

The $Fermi$ LAT Collaboration acknowledges generous ongoing support from a number 
of agencies and institutes that have supported both the development and the 
operation of the LAT as well as scientific data analysis.  These include the 
National Aeronautics and Space Administration and the Department of Energy 
in the United States, the Commissariat \`a l'Energie Atomique and the Centre 
National de la Recherche Scientifique / Institut National de Physique Nucl\'eaire 
et de Physique des Particules in France, the Agenzia Spaziale Italiana and the 
Istituto Nazionale di Fisica Nucleare in Italy, the Ministry of Education, 
Culture, Sports, Science and Technology (MEXT), High Energy Accelerator Research 
Organization (KEK) and Japan Aerospace Exploration Agency (JAXA) in Japan, and 
the K.~A.~Wallenberg Foundation, the Swedish Research Council and the Swedish 
National Space Board in Sweden.

Additional support for science analysis during the operations phase from the 
following agencies is also gratefully acknowledged: the Istituto Nazionale di 
Astrofisica in Italy and the K.~A.~Wallenberg Foundation in Sweden for providing 
a grant in support of a Royal Swedish Academy of Sciences Research fellowship for JC.

This work made extensive use of the ATNF pulsar  catalogue \citep{ATNFcatalog}\footnote{http://www.atnf.csiro.au/research/pulsar/psrcat}.

The LAT team extends thanks to the anonymous referee who made many valuable suggestions of ways to improve this paper. 



{\it Facilities:} \facility{Fermi LAT}.




\bibliography{Bright_Sources_revised5}

\begin{deluxetable}{lrrrrrrrrrrrrclclc}
\setlength{\tabcolsep}{0.04in}
\tablewidth{660pt}
\tabletypesize{\scriptsize}
\rotate
\tablecaption{LAT Bright Source List\label{table1}}
\tablehead{
\colhead{} & \colhead{}  & \colhead{}  & \colhead{}  & \colhead{}  & \colhead{}  & \colhead{} & \multicolumn{3}{c}{100 MeV -- 1 GeV} & \multicolumn{3}{c}{1 GeV -- 100 GeV} & \colhead{} & \colhead{} & \colhead{} & \colhead{} \\ \cline{8-10} \cline{11-13} \\
\colhead{Name 0FGL} &
\colhead{R.A.} &
\colhead{Decl.} &
\colhead{$l$} &
\colhead{$b$} &
\colhead{$\theta_{\rm 95}$} &
\colhead{$\sqrt{TS}$} &
\colhead{$F_{\rm 23}$} &
\colhead{$\Delta F_{\rm 23}$} &
\colhead{$\sqrt{TS_{\rm 23}}$} &
\colhead{$F_{\rm 35}$} &
\colhead{$\Delta F_{\rm 35}$} &
\colhead{$\sqrt{TS_{\rm 35}}$} &
\colhead{Var.} &
\colhead{$\gamma$-ray Assoc.} &
\colhead{Class} &
\colhead{ID or Assoc.} &
\colhead{Ref.}
}
\startdata
 J0007.4+7303 &   1.852 & 73.065 & 119.690 & 10.471 & 0.054 &     64.6 &   32.4 &   1.3 &     36.2 &   6.14 &   0.27 &     55.9 & \nodata & 3EG J0010+7309           & PSR & LAT PSR J0007+7303 & 1 \\
  &  &  &  &  &  &  &  &  &  &  &  &  &  & EGR J0008+7308        &  &  & \\
  &  &  &  &  &  &  &  &  &  &  &  &  &  & 1AGL J0006+7311       &  &  & \\
 J0017.4$-$0503 &   4.358 & $-$5.054 & 101.273 & $-$66.485 & 0.252 &     14.7 &   11.8 &   1.4 &     15.3 &   0.27 &   0.07 &      7.1 & T & \nodata & bzq & CGRaBS J0017$-$0512 & \nodata \\
 J0025.1$-$7202 &   6.295 & $-$72.042 & 305.786 & $-$44.940 & 0.163 &     12.7 &    5.1 &   2.0 &      6.8 &   0.56 &   0.10 &     11.5 & \nodata & \nodata & glb & NGC 104 & 2 \\
  &  &  &  &  &  &  &  &  &  &  &  &  &  &  &  & 47 Tuc & \\
 J0030.3+0450 &   7.600 & 4.848 & 113.111 & $-$57.622 & 0.138 &     18.7 &    8.9 &   0.0 &     13.6 &   0.71 &   0.10 &     14.9 & \nodata & \nodata & PSR & PSR J0030+0451 & 3, 4 \\
 J0033.6$-$1921 &   8.401 & $-$19.360 &  94.215 & $-$81.220 & 0.147 &     10.7 &    1.4 &   0.0 &      4.6 &   0.36 &   0.07 &     10.1 & \nodata & \nodata & bzb & BZB J0033$-$1921 & \nodata \\
\enddata
\tablerefs{{1 \citep{CTA1}, "The Fermi Gamma-Ray Space Telescope Discovers the Pulsar in the Young Galactic Supernova Remnant CTA 1"}, {2 \citep{LAT47Tuc}, "Discovery of high-energy gamma-ray emission from the globular cluster 47 Tucanae with $Fermi$"}, {3 \citep{PSR0030}, "Pulsed Gamma-rays from the millisecond pulsar J0030+0451 with the Fermi Large Area Telescope"}, {4 \citep{LATMSPULSARS}, "Discovery of a Population of Gamma-ray Millisecond Pulsars with the Fermi Large Area Telescope''}, {5 \citep{LSI+61}, "Fermi LAT Observations of LS I +61 303"}, {6 \citep{LATNGC1275}, "Fermi Discovery of gamma-ray emission from NGC1275"}, {7 \citep{LATBlindSearch}, "Sixteen Gamma-Ray Pulsars Discovered in Blind Frequency Searches Using the Fermi LAT"}, {8 \citep{LATVELA}, "$Fermi$ LAT Observations of the Vela Pulsar"}, {9 \citep{LATJ0948}, "The peculiar quasar PMN J0948+0022"}, {10 \citep{PSR1028}, "Discovery of pulsed gamma-rays from the young radio pulsar PSR J1028-5819 with the Fermi Large Area Telescope"}, {11 \citep{LATPSR1048}, "Fermi LAT detection of pulsed gamma-rays from the Vela-like pulsars PSR J1048-5832 and PSR J2229+6114"}, {12 \citep{PKS1454}, "Fermi/LAT discovery of gamma-ray emission from the flat-spectrum radio quasar PKS 1454--354"}, {13 \citep{LATPKS1502}, "PKS 1502+106: a new and distant gamma-ray blazar in outburst discovered by the Fermi Large Area Telescope"}, {14 \citep{PSR2021LAT}, "Pulsed gamma-rays from PSR J2021+3651 with the Fermi Large Area Telescope"}, {15 \citep{LAT2155}, "Resolving the blazar high-energy spectrum of PKS 2155-304 with HESS and FERMI"}, {16 \citep{454.3}, "Early Fermi Gamma-ray Space Telescope observations of the blazar 3C 454.3"}}
\tablecomments{Flux units $10^{-8}$ cm$^{-2}$ s$^{-1}$}
\tablecomments{$\dagger$ - possible SNR or PWN association.  See Table 2.}
\tablecomments{A ``0'' in the $\Delta F_{23}$ column indicates that the entry in the $F_{23}$ flux column is a 2-$\sigma$ upper limit.} 
\end{deluxetable}



\end{document}